\documentclass[%
 reprint,
 amsmath,amssymb,
 aps,
]{revtex4-2}

\usepackage{graphicx}% Include figure files
\usepackage{subcaption}
\usepackage{dcolumn}% Align table columns on decimal point
\usepackage{tabularx}
\usepackage{multirow}
\usepackage{bm}% bold math
\usepackage{amsfonts,color} % Color text for comments
\usepackage[table,xcdraw]{xcolor}
\usepackage[english]{babel}
\usepackage{lipsum} 
\usepackage{array}
\usepackage{float}

\definecolor{Gray}{gray}{0.9}

\newcolumntype{Y}{>{\centering\arraybackslash}X}
\newcolumntype{Z}{>{\raggedright\arraybackslash}X}

\newcolumntype{P}[1]{>{\raggedright\arraybackslash}p{#1}}

\newcommand{\PreserveBackslash}[1]{\let\temp=\\#1\let\\=\temp}
\newcolumntype{C}[1]{>{\PreserveBackslash\centering}p{#1}}

\begin{document}

\preprint{APS/123-QED}

\title{
Modeling when and how physics PhD students search for a research group: the role of interests and prior research experiences in timely group integration
}

\author{Mike Verostek}
  \email{mveroste@ur.rochester.edu}
\affiliation{
 Department of Physics and Astronomy, University of Rochester, Rochester, New York 14627 
}
 \affiliation{School of Physics and Astronomy, Rochester Institute of Technology, Rochester, New York 14623}

 \author{Casey W. Miller}
 \affiliation{School of Chemistry and Materials Science, Rochester Institute of Technology, Rochester, New York 14623}

 \author{Benjamin M. Zwickl}
 \affiliation{School of Physics and Astronomy, Rochester Institute of Technology, Rochester, New York 14623}

\date{\today}

\begin{abstract}

Studying the factors that influence the quality of physics PhD students' doctoral experiences, especially those that motivate them to stay or leave their programs, is critical for providing them with more holistic and equitable support.  Prior literature on doctoral attrition has found that students with clear research interests who establish an advisor-advisee relationship early in their graduate careers are most likely to persist.  However, these trends have not been investigated in the context of physics, and the underlying reasons for why these characteristics are associated with leaving remain unstudied.  Using semi-structured interviews with 40 first and second year physics PhD students, we construct a model describing the characteristic pathways that physics PhD students take while evaluating interest congruence of prospective research groups.  We show how access to undergraduate research and other formative experiences helped some students narrow their interests and look for research groups before arriving to graduate school.  In turn, these students reported fewer difficulties finding a group than students whose search for an advisor took place during the first year of their PhD.  Lastly, we identify two characteristic types of students at a higher risk of leaving their programs: students who enter graduate school with broad interests and struggle to find a group, and students who join a research group early based on research interest alone and subsequently encounter issues with a negative mentoring relationship.  This work serves as a major step toward creating a comprehensive model of how PhD students find a research group, and opens the door for future work to investigate how factors such as group culture and working environment impact the search process.   

\end{abstract}

\maketitle
%\tableofcontents

\section{\label{sec:Introduction}Introduction and Background}

Graduate school is known to be a complex and intense environment for first-year PhD students.  Adapting to the norms and expectations of a new and unfamiliar department can be difficult, as is the process of transitioning into the role of professional researcher.  During this critical period, finding the right research advisor is widely acknowledged to be one of the most impactful events for students on the path to a PhD \cite{lee2008doctoral, rigler2017agency}.  Finding an effective research mentor and group provides the basis for students to build specialized research skills \cite{juliano2001critical, feldon2019postdocs}, embrace their identities as scientists \cite{malone2009narrations, lane2019model}, and develop a network of support among their labmates \cite{griffin2018supporting, stachl2020sense}.  On the other hand, poor advising relationships negatively affect PhD students' intellectual development, mental health, and willingness to pursue an academic career after graduation  \cite{cohen2022abuse, martin2013countering, gin2021phdepression}.  For some students, a negative advising relationship is one of the main factors that motivates them to leave their programs  \cite{lovitts2002leaving, devos2017doctoral, bair2004doctoral, rigler2017agency, golde1998beginning, jacks1983abcs}.  

% Significant prior literature across STEM and non-STEM disciplines has gone toward identifying the characteristics indicative of productive mentorship  \cite{schlosser2003qualitative, schlosser2011multiculturally, barnes2010characteristics, bargar1983advisor}.  However, less work has examined the process by which students go about finding a research group in the first place.

Studying the factors that influence the quality of PhD students' experiences and motivate them to stay or leave their programs is an important goal for researchers across graduate education \cite{lott2009doctoral, sachmpazidi2021departmental, maher2020exploring, wilson2012hierarchical, miller2019typical}.  Better understanding why doctoral students leave is particularly critical for creating a more diverse and equitable graduate landscape, as attrition from PhD programs disproportionately affects traditionally underrepresented students \cite{cgs2015minorityphd, fisher2019structure, cidlinska2023don}.  In the context of physics, the retention rate of PhD students is estimated to be approximately 50\% \cite{cgs2008phd, cgs2015minorityphd}.  For students, leaving a PhD program may mean grappling with feelings of failure and disappointment, which can have adverse effects on their emotional and physical well-being \cite{lovitts2002leaving}. Graduate departments are not left unscathed by high rates of attrition either, as they must invest future resources into recruiting and supporting new students \cite{golde2005role}.

Most attrition from physics PhD programs occurs in the first two years \cite{verostek2021analyzing}, but research shows that students begin to think about leaving long before they actually do so.   Students who leave after their first year of graduate school typically begin contemplating their decision to leave by their second semester \cite{lovitts2002leaving}.  This timeframe coincides with when many students are likely in the midst of attempting to find a research group, and their success in that process may impact their decision to stay or leave.  However, tracking when students join and leave research groups is not a common metric for departments to record, which leaves it unclear when students actually take this critical step in their graduate careers.  Hidden inequities are most likely to go unnoticed in graduate education when student outcomes and progress are poorly tracked \cite{lovitts2002leaving}; thus, better understanding the factors that influence when students join a research group may yield insight into systemic inequities inherent in the group search process.  

In a seminal multidisciplinary study of doctoral attrition, \citeauthor{lovitts2002leaving} found that a combination of students' research interests, proactivity in seeking out a research group, and closeness of interaction with prospective advisors prior to graduate school were all predictive of students' likelihood of PhD completion.  She noted that ``few noncompleters enter graduate school knowing who they want to work with,'' and that ``having interests that are close to a faculty member's \textit{and} establishing the advisor-advisee relationship early in the graduate career based on firsthand knowledge and interaction'' contribute most to successful PhD completion.  Thus, \citeauthor{lovitts2002leaving} contends that students with strong research interests who search for and join a group early are most likely to persist.  

However, the underlying reasons driving these trends remain unclear.  For example, it is not obvious that having broad research interests should correlate with a lower likelihood of PhD completion; on the contrary, intellectual curiosity and a willingness to explore new topics are traits often valued in science.  Furthermore, physics was one discipline omitted in \citeauthor{lovitts2002leaving}' study, leaving it unknown whether these trends apply to physics students as well.  To fill this gap, this study seeks to characterize physics PhD students' research interests as they enter graduate school, and to better understand the influence that these interests have on their search for a research group.  In doing so, we also seek to better understand how these previously identified predictors of attrition (strength of interest, time students look for and join research groups) are linked, as well as uncover the underlying factors driving these trends. 

Little is known about the extent to which physics PhD students know what research they want to pursue upon entering graduate school.  Studies by \citeauthor{maher2020finding} in the context of biology PhD programs have shown that biology graduate students place a high emphasis on finding a group with congruent research interests, despite often being unsure of exactly what research topic they were interested in pursuing.  Biology PhD students commonly hoped to explore their interests while in graduate school \cite{maher2019doctoral, maher2020finding}, and interest exploration in biology is encouraged through highly formalized lab rotation systems.  The authors found these rotations benefited students in trying to narrow their interests.  Such formal requirements are comparatively rare in physics graduate programs though \cite{feder2020phd, artiles2023doctoral}.  Rather, as detailed in our previous work, physics PhD students often sought more structure for meeting groups, which may harm students with broader interests if they encounter more difficulties finding the right fit \cite{verostek2023physics}.  

That study showed other difficulties that physics students described as well, including trouble understanding the research that prospective groups were doing, struggling to connect with faculty and graduate students, and lacking time to search due to coursework.  Moreover, students who struggled to find a group tended to report a lower sense of belonging in their programs during their first year, which led several of them to consider leaving their programs.  On the other hand, successfully joining a group often granted students a greater sense of community and purpose in their programs that motivated them to stay.  These results are supported by prior research that has shown how becoming a part of a research group can give STEM students opportunities to develop their identities as scientists \cite{malone2009narrations, graham2013increasing} while increasing their sense of belonging to their research community \cite{o2017sense, stachl2020sense,alaee2022impact, dolan2009toward, lopatto2007undergraduate}.  

Students who are able to join a research group early in their graduate careers may enjoy a variety of other benefits as well.  For instance, entering into a group early may accelerate students' time to doctoral completion \cite{verostek2021time}, which is known to positively correlate with a student's salary upon graduation \cite{potvin2012examining}.  By finding a group sooner, students might experience less anxiety about needing to find a group in the midst of classes, and concerns about funding and competition for spots might be assuaged.  Indeed, the number of research groups available for students to join in a department is always limited by funding and space, which naturally generates competition for this finite resource \cite{mendoza2007academic, slaughter1997academic, stephan2012economics}.

Yet access to a research group, and all of the benefits that are known accompany it, may strongly depend on students' research interests and how efficiently they are able to navigate the search process.  Students whose backgrounds offer them more opportunities to narrow their research interests and make connections with graduate faculty before entering graduate school may be better positioned to more quickly begin research.  Research on the impacts of undergraduate research experiences has shown that students tend to constrain their options for future work to topics with which they are familiar \cite{laursen2010undergraduate, national2017undergraduate}.  Indeed, ample research has illustrated that physics students' interests are most informed by their backgrounds and learning experiences \cite{bandura1982self, lent2000contextual, bennett2022analysis}, and access to undergraduate research experiences are particularly critical \cite{alaee2022impact, lopatto2007undergraduate, cardona2021access}. 

Despite the fact that undergraduate research experiences are often critical for students' interest development, access to undergraduate research is lower for traditionally underrepresented groups in physics, particularly ethnic minorities and first-generation college students \cite{kuh2008high}.  In another previous work, we revealed how access to undergraduate research can have significant repercussions during the group search \cite{verostek2023inequities}.  We identified a number of advantages that one student enjoyed while navigating the search process: attending an elite undergraduate institution provided a leg up in admissions, an REU at his top-choice institution gave insight into which labs he might want to join, and advice from his undergraduate advisor helped him evaluate prospective advisors.  Meanwhile, another student described a number of difficulties navigating graduate school, many of which she attributed to a lack of departmental support for first-generation students. She expected significantly more academic advising from her graduate school to help her through her first year and struggled to find a research lab, which negatively influenced her sense of belonging in the program.  Collectively, prior literature suggests relationships between students' background experiences, their research interests, and when and how they go about searching for a research group in graduate school.

In this study we seek to better understand how physics PhD students find a group with congruent research interests.  While we acknowledge that other characteristics (e.g., the group's culture, work-life balance) are also important to students, the present work focuses on the role of research interests and was guided by the following questions:

\begin{enumerate}
    \item What factors (both before and during graduate school) influence when physics PhD students join a research group?
    \item Specifically, what role does research interest play in guiding students' search?
    \item How do the concerns and difficulties that students experience during their search for a group differ based on when students look for a group?
    \item What are the consequences of these differences on students' success in their PhD programs?
\end{enumerate}

Sections \ref{subsec:whenjoin} to \ref{subsec:stages} attend to answering RQ1 and RQ2.  In Section \ref{subsec:whenjoin}, we outline when students in our sample found their research groups and identify a subset of students in our sample who committed to a research group early in their PhD timelines.  We also illustrate a number of benefits that these students associated with doing so. Section \ref{subsec:interests} demonstrates that physics PhD students enter graduate school with drastically different levels of certainty regarding which research they wanted to pursue, and Section \ref{subsec:intereststime} discusses how this correlates with when students join research groups.  Then, Section \ref{subsec:stages} investigates when students look for research groups, showing how students who engage in the search process earlier tend to join a group more quickly.  Throughout each of these sections, we provide evidence to highlight how students' access to resources and experiences during their undergraduate studies impacted their search process.  In Section \ref{subsec:concerns}, we address RQ3 by showing how the frequency and types of concerns and difficulties that students experience during their search process vary depending on when they joined a research group.  Lastly, Section \ref{subsec:leavingsubset} focuses on RQ4 by closely examining the cases of several students who considered leaving their PhD programs, culminating in a model that identifies two characteristic types of students at a higher risk of leaving.

\begin{figure}[t]
\centering
\includegraphics[]{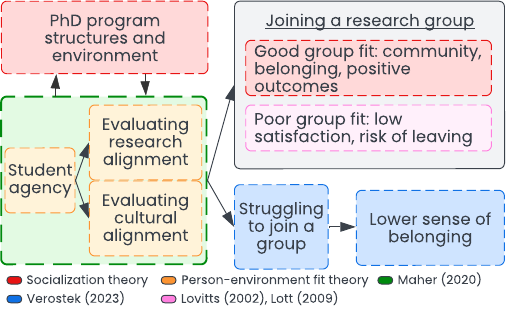}
\caption{\label{fig:intro} A framework illustrating how we are conceptualizing the elements of the group search process in physics.  Colors illustrate the prior literature or theory that motivated the inclusion of each element.}
\end{figure}
\section{\label{sec:theory}Conceptual Framing}

We leveraged insights from two theoretical frameworks while designing the current study, following the example of \citeauthor{maher2020finding} \cite{maher2020finding}.  This section briefly describes these frameworks and how they informed our methodological approach (see Figure \ref{fig:intro}).  First, we discuss \textit{person-environment fit}, which has recently been suggested as a useful theoretical lens through which to examine doctoral education \cite{baker2015antecedents}.  We then discuss \textit{socialization theory}, which is the most prevalent framework used in studies of US doctoral education \cite{austin2006preparing}.  

With historical origins in vocational selection \cite{parsons1909choosing, pervin1968performance, holland1959theory, holland1997making}, person-environment fit research focuses on the compatibility of individuals within their working environments \cite{brown2002career, van2016study, osipow1990convergence, su2015person}.  The overall construct of person-environment fit consists of multiple dimensions of compatibility, including person–vocation (compatibility between an individual and overall career path), person–job (more narrowly defined between a person and its associated responsibilities), person–organization (an individual and their overall organization), person–group (a person and the individuals in their work group), and person–person (a person and their mentor or supervisor) \cite{jansen2006toward}.  The theory therefore differentiates between aspects of fit more closely corresponding to one's profession and its associated tasks (vocation, job), and those more closely associated with interpersonal concerns (group, person).  This distinction further motivated our decision in this study to narrow the focus of this study to specifically examine the role of students' research interests on their group search.

Three primary assumptions underlie person-environment fit: 1) people seek out environments that match their personality and abilities, 2) person-environment fit is a reciprocal process in which people can shape their environment and the environment can influence people, and 3) work outcomes (e.g., satisfaction, performance) are dependent on the interactions between individuals and their environments, with congruent fits resulting in positive outcomes \cite{su2015person}.  For the purposes of this study, we view a person's \textit{perceptions} about an environment as the most important indicator of fit.  Although level of congruence between and their environment may be induced more ``objectively'' by measuring and assessing individual and environmental variables \cite{dawis2002person, kristof2012organizational, su2015person}, perceived fit highlights the way individuals feel about aspects of their environment and allows people to give varying levels of salience to different features.

Meanwhile, socialization theory describes the processes through which doctoral students internalize disciplinary values, norms, expectations, and knowledge via formal and informal interactions with advisors, faculty, and peers  \cite{weidman2001socialization, weidman2020toward, gardner2008fitting, austin2006preparing}. These processes are often characterized as occurring over a series of stages \cite{thornton1975dynamics}.  The first stage is anticipatory, as students enter the program with preconceived notions of what the PhD entails and begin to meet with faculty and students.  Formal socialization processes ensue, including coursework and beginning research in a lab \cite{weidman2020toward}.  For students in the sciences, participation in a research group is a primary source of socialization, as it allows for development research skills and affords them access to a network of faculty and peers \cite{gardner2007heard,gopaul2011distinction}.  This aligns with our prior results suggesting that students reap a variety of benefits from becoming integrated into research early in their PhD timelines.

The final stage of socialization is internalizing the role of an independent scholar and integrating this new identity into existing ones.  Socialization frameworks have been used as a lens to examine attrition from doctoral programs, and are particularly useful for conceptualizing the dual nature of PhD students as both students and professionals.  \citeauthor{golde1998beginning} refers to this as an ``unusual double socialization'' of graduate students \cite{golde1998beginning}, who in addition to integrating into their individual department in their role as students must also begin to operate as professional physicists in their role as researchers \cite{gardner2010contrasting}.  Along both of these axes, faculty clearly play a critical role as both instructors and advisors.  Critically, poor socialization has been linked to students' decisions to leave graduate school \cite{lovitts2002leaving, golde2000should}.  These findings highlight how inadequate structural support for students, rather than perceived shortcomings in individual students' behaviors and characteristics, can lead to early departure.  However, socialization has sometimes been criticized for downplaying the role of student agency as they move through their programs \cite{tierney1993enhancing, gopaul2011distinction}, leading to more recent iterations of the framework to more explicitly include the student's role in advancing their own socialization \cite{portnoi2015expanding, weidman2020toward}.  

Both person-environment fit and socialization provide useful theoretical lenses through which to conceptualize physics PhD students' search for a research group.  The core assumptions of person-environment fit theory indicate that students will actively seek to match their interests and values to those they perceive in their prospective research group. Hence, in the absence of formal departmental structures such as lab rotations to facilitate access to groups, we view student agency having a significant role to play in physics PhD students' search for a research lab.  Meanwhile, in the context of socialization theory, successfully navigating the group search process is the antecedent to accessing the numerous and impactful socialization opportunities associated with doing research.  Thus, student agency can grant access to socialization experiences.  We also expect this relationship to be reciprocal.  For example, socialization experiences during the first year may provide students with knowledge that allows them to more easily navigate the search process.  However, we also recognize that the group search process is embedded within the structural constraints imposed by each student's graduate program.  Hence, we also wish to highlight the ways students feel inhibited in their search.

%%%%%%%%%%%%%%%%%%%%%%%%%%%%%%%%%%%%%%%%%%%%%%

\begin{table*}[t]
\renewcommand\tabularxcolumn[1]{m{#1}}
\centering
\def\arraystretch{1.1}%  1 is the default
\begin{tabularx}{\textwidth}{XYYYYYYY}
\cline{1-8}
 &White/ caucasian & Hispanic, Latinx, or Spanish origin& Black or African American & Asian & North African & From multiple races & \textbf{Total} \\ \hline \hline
 
\multicolumn{1}{l}{Male} & 6 & 4 & 0 & 5 & 0 & 0 & \textbf{15} \\ 
\rowcolor[gray]{.9}[\tabcolsep] 
\multicolumn{1}{l}{Female} & 10 & 3 & 2 & 4 & 1 & 2 & \textbf{22} \\ 
\multicolumn{1}{l}{Non-binary} & 2 & 1 & 0 & 0 & 0 & 0 & \textbf{3} \\ 
\rowcolor[gray]{.9}[\tabcolsep] 
\multicolumn{1}{l}{\textbf{Total}} & \textbf{18} & \textbf{8} & \textbf{2} & \textbf{9} & \textbf{1} & \textbf{2} & \textbf{40} \\ 
\hline
\end{tabularx}
\caption{\label{tab:demographics1} Gender and race data for the $N=40$ interviewees in the sample Demographic information was gathered using a Qualtrics survey prior to each interview.  Surveys were fixed-choice, but contained a space for respondents to input answers outside the choices provided.  Fixed-choice responses for race/ethnicity questions were drawn from the US Census American Community Survey \cite{census}.}
\end{table*}

\section{\label{sec:Method}Method}

\subsection{Methodological approach and interview protocol}
We sought out methodological approaches that would allow us to explore the role of student agency as well as structural constraints in the group search process.  We drew inspiration from cognitive task analysis (CTA) methods \cite{crandall2006working, clark2008early} and Dervin's sense-making method \cite{dervin1986neutral, dervin2003sensemaking}. According to \citeauthor{crandall2006working} in their handbook on cognitive task analysis, CTA methods are meant to ``understand how people think: how their minds work, what they struggle with, and how they manage to perform complex work adeptly'' \cite{crandall2006working}.  For example, analogous to the present work on how students ``shop'' for a research advisor, such methods have been used to study how consumers gather information about products and choose which ones to purchase  \cite{reinhard2012comparing}.  One critical aspect of understanding the cognitive processes underlying individual behavior is how individuals evaluate and respond to their external environment \cite{klein2010rapid, fackler2009critical, gazarian2010nurse}.  Similarly, sense-making operates under the assumption that humans move in and out of structures that influence their behavior, sometimes making sense of them, resisting them, or creating new structures altogether \cite{dervin2015dervin}.  Both methodologies are designed to elicit detailed descriptions of interviewees' thoughts and actions as they recount how they performed a task, focusing on understanding their specific life experiences in great depth.

Our semi-structured interview protocol asked students to construct a timeline of steps they took while searching for a research group (Stage 1).  Students were free to start their timeline at any point, but most commonly they began during their junior or senior years of undergraduate study.  Timelines often consisted of four or five main steps.  For each step, we asked students about any major questions and concerns they had at each step of their timeline (Stage 2).  This included anything they wanted to find out, were confused about, worried about, or were just curious about.  Many responses to this part of the protocol revolved around students' internal deliberations regarding what research they wanted to pursue in graduate school, as well as their efforts to gather information about prospective groups' research.  Students also often commented on general difficulties they experienced during their search, as well as concerns about their future in graduate school.  Lastly, we asked students to identify any sources of help that allowed them to resolve their question or concern, as well as any obstacles that hurt their ability to move forward (Stage 3).  This third stage allowed us to understand the thoughts, actions, and events that helped or hurt students' ability to navigate their search for a group.  The analysis presented here primarily focuses on several of the major questions and concerns that students discussed in Stage 2, but some supporting evidence was drawn from Stage 3 of the protocol focusing on sources of helps and hurts.  The full interview protocol used in this study is available in the Supplemental Material \cite{supplementalMats}. 

We also leveraged the data from Stage 1 to identify when each event occurred in their academic timeline.  Students typically described when events occurred in terms of academic periods.  Using this information, we were able to identify each excerpt as occurring at one of seven points in time: Before Senior year, Senior year, Senior summer, 1st semester, 2nd semester, 1st year summer, and 2nd year.  Senior year refers to students' final year of undergraduate study, whereas 1st and 2nd semester refer to the first and second semesters in their graduate program.  We use this language in the results section \ref{sec:Results} to indicate when in the student's timeline they said it.

\subsection{Recruitment and participant sample}

This study is part of an ongoing analysis aimed at systematically characterizing the process by which PhD physics students search for a research group.  Study participants were recruited by emailing physics graduate program directors and asking them to forward our recruitment letter to their first and second year graduate students.  A \$25 Amazon gift card was offered as incentive for participation in our study.  We targeted these years of study because they were either in the process of or had recently completed searching for a research group.  We also intentionally chose to email programs of varying size and research activity to ensure a variety of institutional contexts were represented.  In total, we reached out to 18 graduate programs directly.  Ten programs forwarded our recruitment email to their first and second year graduate students.  Since we reached out to programs rather than individual students, we do not know precisely how many students received invitations to participate.  However, physics graduate programs admit an average of 16 students per year \cite{aip2019firstyear}, meaning that the number of students invited to be interviewed was likely on the order of several hundred participants.  A more detailed breakdown of institution types and their program requirements is available in the Supplemental Material \cite{supplementalMats}.

Since this project is conducted with the support of the Inclusive Graduate Education Network (IGEN), a partner of the American Physical Society (APS), one major goal of this work is to improve diversity and inclusion across physics graduate education.  One of the leading efforts toward that end is the APS Bridge Program, which is a post-baccalaureate program designed to increase the number of PhDs earned by underrepresented students in physics \cite{aps_bridge_home}.  Hence, we also sent our recruitment information directly to APS Bridge students currently in their PhD programs to help ensure our data represented their experiences as well.

The sample of students in this analysis consists of 40 students representing 13 institutions; 5 students were from the Bridge program, all from different institutions.  A fixed-choice demographic survey was administered prior to each interview via Qualtrics, but each response field contained a space for respondents to input answers outside the choices provided.  Based on the results of the survey, $N=22$ interviewees identified as women, $N=15$ identified as men, and $N=3$ identified as non-binary or gender fluid.  $N=21$ were in their first year of physics graduate school while $N=19$ were in their second year.  Fixed-choice responses for race/ethnicity questions were drawn from the US Census American Community Survey \cite{census};  $N=18$ identified as White/Caucasian, $N=9$ as Asian, $N=8$ as Hispanic/Latinx or Spanish origin, $N=2$ as Black or African American, $N=1$ as North African, and $N=2$ from multiple races.  $N=32$ students were from the US while $N=8$ were non-US students.  Table \ref{tab:demographics1} offers a more detailed demographic breakdown.  Interviews were conducted over Zoom and took place from Fall 2022 to Spring 2023.  All names used throughout the paper are pseudonyms.

\subsection{Data analysis}

Once completed, interviews were transcribed and edited for grammar and clarity.  The transcripts became the subject of our thematic analysis, which followed the steps outlined in \citeauthor{braun2006using} \cite{braun2006using}. Analysis began with repeated and active reading of the data and application of a priori codes aligned with the stages of the protocol (e.g., ``Step'' or ``Question'').  For each, we also applied an initial code describing the excerpt and what we found important about it, which is a common first step in grounded theory approaches to data analysis \cite{charmaz2006constructing}.  Initial coding is an open-ended first cycle coding process that involves breaking down qualitative data into discrete parts, examining them closely, and applying codes that promote deep reflection on the contents of the data \cite{saldana2013introduction}.  Our initial codes typically consisted of a short phrase or sentence describing the excerpt and what we found important about it.  

After applying this process to 20 of the interviews, we started to focus on inductively identifying emergent codes to capture interesting features about questions and concerns that students had.  We reduced the types of questions into three primary themes: \textit{Research alignment}, \textit{Group culture}, and a third theme of \textit{Overall concerns} about the search process that did not fit neatly fit into either of the other two.  The present study primarily focuses on the \textit{Research alignment} theme, which contained statements regarding students' efforts to find a research group that aligns with their interests.  We also provide more nuanced descriptions of several sub-themes from the \textit{Overall concerns}, which were detailed in an earlier analysis using data from 20 of the 40 interviews \cite{verostek2023physics}.  

At this point we developed a codebook specifically focused on the research alignment aspect of students' narratives.  We coded excerpts along two broad axes: comments regarding students' personal \textit{Interest development}, and statements about how students characterized the process of \textit{Evaluating research alignment}.  Excerpts coded under \textit{Interest development} were indicative of how well a student knew what research they wanted to pursue in graduate school.  The code \textit{Evaluating research alignment} was applied to comments in which students described the characteristics they were looking for in a research group with regard to congruence of research interests.  In some cases, both codes were applied to an excerpt.  For instance, the excerpt ``I was looking at programs to apply for... and if I couldn't find people that did theoretical stellar astrophysics I crossed them off the list'' was double coded to capture the interest shown in theoretical astrophysics and the activity of evaluating programs' available research.  As the coding process continued, these codes were refined and broken into new themes and sub-themes.  The results of our analysis are presented in Section \ref{sec:Results}.  In particular, results stemming from our analysis of student \textit{Interest development} are presented in Section \ref{subsec:interests} and results from our analysis of student efforts when \textit{Evaluating research alignment} are presented in \ref{subsec:stages}.

In addition to thematically organizing the data, we also linked student-level meta data information to all 40 transcripts regarding when each student committed to joining their current research group.  While many students had a clear idea of when they committed to their research group (e.g., ``I shook on it with my PI, saying I was going to commit to joining the lab fully.''), this was not always the case.  For instance, some students described having an interest in a group before graduate school and beginning to perform research activities with the group in their first semester, but never described a distinct moment when they joined the group.  Approximately $N=7$ students presented this ambiguity.  The authors (M.V.) and (B.Z.) discussed these cases individually. Together they agreed on the time when these students most strongly began to characterize themselves as members of their group, typically evidenced by participation in group research activities and statements about their confidence in which research topic they wanted to pursue.  For example, one student said that after she met her current advisor during her senior summer, ``I don't think I ever really thought to look for a different advisor.''  Still, we classified this student as joining her group in her first semester of graduate school when she actually began to perform research activities in this advisor's group.  Our analysis of when the physics PhD students in our sample joined their research groups is discussed in Section \ref{subsec:whenjoin}.  

We used several strategies to produce a valid and reliable coding scheme.  One researcher (M.V.) was responsible for developing most of the codes and themes, which were refined through discussion between (M.V.) and (B.Z.) at weekly research meetings.  Inter-rater reliability was done once the final codebook was established in order to demonstrate its validity.  In particular, a selection of 40 random excerpts corresponding to students' extent of interest development and stage of evaluating research were given to a researcher unaffiliated with the project to code using the sub-themes discussed in the results.  Agreement was high, with disagreements primarily occurring over five borderline cases (e.g., whether a student who had primarily looked at only a few research groups before arriving to graduate school, but had never contacted them, would count as evaluating school level research alignment or evaluating individual group level alignment).  We refined the codebook after discussion better delineate when certain codes should be applied.

%\subsection{Analysis of program handbooks}

\begin{figure*}[t]
\centering
\includegraphics[]{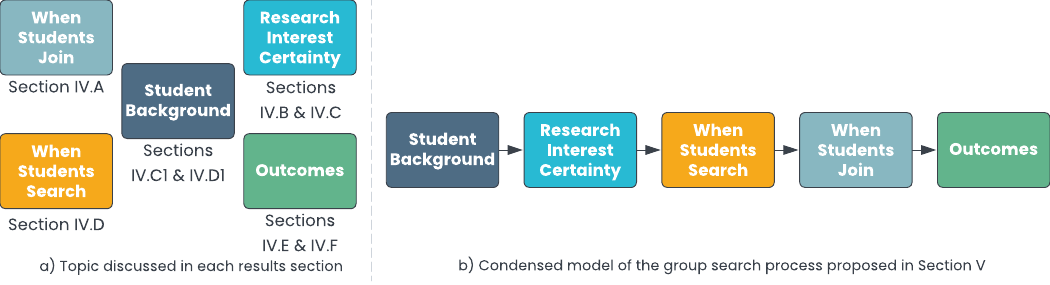}
\caption{\label{fig:roadmap} (a) Summary of the main topics discussed in the results section.  Although the impact of students' background experiences on the group search is mentioned throughout the results, it is the focus of subsections \ref{subsubsec:ugresearch} and \ref{subsubsec:ugconnections}. (b) Condensed version of the model presented in Section \ref{sec:Discussion}, previewing the relationships between the variables that we establish throughout the results.}
\end{figure*}

Lastly, to gain more information about the institutional requirements regarding when students should (or must) join a group, we analyzed the graduate program handbooks of the schools represented in the sample.  Only 2 of the 13 schools did not have an online graduate student handbook available.  One school did not appear to have any publicly available resources for expected PhD timelines and program requirements.  For analysis, we examined the resources to find the first mention of students needing an advisor.  We collected these instances and grouped them based on whether finding an advisor was an explicit or implicit requirement, as well as the specificity of when in students' PhD timeline those requirements occurred.  Results are presented in Section \ref{subsec:whenjoin}.  An overview of the results section and which major topics are covered in each section is given in Figure \ref{fig:roadmap}.

%%%%%%%%%%%%%%%%%%%%%%%%%%%%%%%%%%%%%%%%%%%%%%

\section{\label{sec:Results}Results}

\subsection{\label{subsec:whenjoin} When physics PhD students join research groups and the perceived benefits of joining sooner}

% Demarcating the first semester or before as an ``early'' time to join a group also aligns with \citeauthor{lovitts2002leaving}' observation that the \textit{second} semester is a critical period during which students begin to consider leaving. 

\citeauthor{lovitts2002leaving} found that PhD students who establish the advisor-advisee relationship early in their graduate careers are more likely to complete their PhD successfully \cite{lovitts2002leaving}.  However, the definition of ``early'' is unclear in the context of physics graduate education, as there is no prior literature documenting when students join research groups.  By examining when students in our sample found their research groups, we motivate our decision to categorize students who joined a group in their first semester or before as ``early joiners.''  These students perceived several benefits from joining a group quickly, and noted that their experiences during the first year contrasted with those of students who committed to a research group later on.  We also motivate our definition of ``joining early'' with an analysis of program handbooks to better understand when departments expect students to find a group.     

\textbf{Programmatic expectations} Analysis of departmental expectations of when students should join a group reveals strongly varying levels of guidance across the programs studied.  Only one provided a specific time by which PhD students were \textit{required} to have found an advisor (the fall semester after passing a written qualifying examination).  On the other hand, 5 schools structured their expected PhD timelines around completion of courses and qualifying exams, and did not explicitly provide a time by which students were expected to join a group.  Rather, joining a research group was implicitly needed in order to complete the other program requirements.  For instance, one school required that students pass an oral exam on a topic jointly chosen by the student and their research advisor by the beginning of the third year.  Finding an advisor is clearly necessary for this step, but it was not an explicit expectation in the timeline.  

Most commonly, programs provided a ``typical timeline'' or an ``expected timeframe'' for finding a research group.  Six schools fell into this category, but their expectations varied.  For example, one strongly suggested students should join a group in the spring semester of their first year, while another suggested that it was more typical to ``formally'' join a research group in the third year of the program.  Another provided a broad swath of time (spring semester of first year to summer of the third year) during which finding an advisor was expected, but noted that earlier was better.  The lone program that employed a formal laboratory rotation requirement for students expected them to have chosen their research advisor by the end of their first academic year.  Despite ambiguity across most of the program handbooks regarding when students are expected to find an advisor, they generally indicate that a ``typical'' timeline for students to find an advisor should be around the end of their first academic year or beginning of their second year.

\begin{table}[b]
\centering
\def\arraystretch{1.1}%  1 is the default
\begin{tabularx}{\columnwidth}{|>{\hsize=1.5\hsize}X|
                               > {\hsize=.5\hsize}Y|}
\hline
\textbf{When committed to group} &  \textbf{N} \\ \hline

\multicolumn{1}{|l|}{Senior year} & 
\multicolumn{1}{c|}{1}     \\ 
 
\multicolumn{1}{|l|}{Senior summer} & 
\multicolumn{1}{c|}{7}    \\ 

\multicolumn{1}{|l|}{1st semester} & 
\multicolumn{1}{c|}{10}    \\ 

\rowcolor[gray]{.9}[\tabcolsep] 
\multicolumn{1}{|l|}{\textit{Subtotal: 1st semester or before }}  & 
\multicolumn{1}{c|}{\textit{18}}    \\ 
\hline

\multicolumn{1}{|l|}{2nd semester}  & 
\multicolumn{1}{c|}{9}     \\ 

\multicolumn{1}{|l|}{1st year summer} &
\multicolumn{1}{c|}{7}    \\ 

\multicolumn{1}{|l|}{2nd year or later} & 
\multicolumn{1}{c|}{3}   \\ 

\multicolumn{1}{|l|}{Has not joined} & 
\multicolumn{1}{c|}{3}   \\ 

\rowcolor[gray]{.9}[\tabcolsep] 
\multicolumn{1}{|l|}{\textit{Subtotal: 2nd semester or after}} & 
\multicolumn{1}{c|}{\textit{22}}    \\ 

\hline
\multicolumn{1}{|l|}{\textbf{Total}}  & 
\multicolumn{1}{c|}{40}  \\ 
\hline
\end{tabularx}
\caption{\label{tab:whenjoin} The semester that students in our sample reported committing to their graduate research group.  $N=18$ students joined in their first semester or before, which implies that much of their search took place before arriving to graduate school.  $N=22$ joined in their second semester or later, aligning with more typical timelines suggested in graduate handbooks.  Three students had not joined a group at the time of their interviews, but these students were either already in their second semester of graduate school or were nearly finished with their first semester.  Thus, we categorized their timelines with those students who joined after their second semester.}
\end{table}

\textbf{When students join research groups} The total number of students in our sample who joined a group during or after their 2nd semester of graduate school was $N=22$.  16 of the 22 students joined before their second year began, which falls within the expected timeframe of most PhD programs.  Meanwhile, three joined in the middle of their second year or later.  Three students had not joined a group at the time of their interviews, but these students were either already in their second semester of graduate school or were nearly finished with their first semester.  Thus, we categorized their timelines with those students who joined after their second semester.    

Nearly half of the physics PhD students in our sample ($N=18$) had committed to a research group during or before their first semester of graduate school (see Table \ref{tab:whenjoin}).  One student committed to a group during her senior year of undergraduate study (she continued her graduate study at her undergraduate institution, and had agreed to work with her undergraduate research advisor before graduate school began), and seven students committed during the summer after their senior year.  The remaining ten committed during their 1st semester of graduate school.

In order to have joined a group in their first semester of graduate school or before, those students likely conducted much of their search before they arrived.  This is particularly true of the students who started working with their advisors the summer before their programs started, who were well ahead of the expected timelines published by their universities.  On the other hand, students who joined in their second semester or later more likely searched for groups during their first year of graduate school, and their timelines align more closely with what program handbooks describe as more ``typical.''

Notably, the time to join a group reported in Table \ref{tab:whenjoin} represents the time that students took to join their \textit{most current} research group.  However, several had joined earlier than the time indicated in the table and subsequently chose to switch.  $N=7$ of the 22 students who reported finding a group later in their graduate programs had done so after switching out of a different group.  For a majority of students, switching research groups was a personal decision associated with a unique set of challenges.  Thus, we chose not to categorize the three students in our sample whose programs had formal rotations as having ``switched.''  Although these students changed research groups during their first year, it was a mandatory part of their program rather than a decision they made on their own.

\textbf{Students distinguish joining earlier or later} Students who joined a research group before or during their first semester discussed several perceived benefits compared to their peers.  Several students noted that getting into a group before the first year of coursework reduced anxiety about their future due to the security it offered them.  Nina expressed how finding a group ``did take a lot of weight off my shoulders... not necessarily having to spend too much time obsessing about oh, everybody else already has a group.''  Luis had two offers for research groups before his first year of graduate school began, which he described as an advantage over other students in his class: ``I felt so relieved that whole first year, that I didn't have to go out and look for a group as actively as everyone else did right?... a lot of people really struggled to find a good fit, whereas I felt like, `Oh, I could send an email right now and get it in the group that I want, or in a group that wants me.'''  In both of these quotes, students describe comparing their own progress to other students in their cohort, which eased their anxiety.  However, the reverse was true for students who did not join a group right away. Dev, who struggled to determine if his preferred advisor would be able to fund him or not, said during his second semester of graduate school ``It is a bit stressful though in the sense that, for some of the people, before even coming here they had fixed on an advisor. Everything was fixed... they'll go and do the PhD. So talking to [those students] was not really helpful because they are not going through any of this.''  Dev's quote highlights feeling like an outsider compared to other students who seemed to be further ahead.

Others noted that getting into research faster made them feel like they were making progress toward their degree.  Heather credited joining a group during her first semester of graduate school for her accelerated graduation timeline: ``if I had not been in [my advisor's] group as a first year and kind of set on that path, I don't think I would have had the resources or knowledge to be ready to [take my qualifying exam] that that early.''  Being able to partake in research during her first year allowed Heather to take her research-focused qualifying exam earlier than most other students in her class.  This then allowed her to apply for several grants that were only available to students who had completed their qualifying exam, furthering her research agenda.  On the other hand, Jack had not joined a group at the time he was interviewed in his second semester.  He was hoping to join a group that he perceived as highly competitive, and he had taken the professor's elective class during his first semester to demonstrate his interest.  Jack was concerned about being behind his desired PhD timeline if he was unable to join his preferred group at the end of his second semester: ``I've put in a lot of time into this, and if it ends up not working out, it's a bit annoying because of, yeah, just wasted time. I'd say it's a bit stressful for sure.''

The comments in this section motivate our choice to demarcate the first semester or before as an ``early'' time to join a group.  They also begin to demonstrate the effect that the time students join a group can have on their emotional state and overall outlook on how well they are progressing through their PhD program.  Those themes broadly align with our prior findings that students who struggled to find a research group tended to feel a lower sense of belonging in their programs \cite{verostek2023physics}.  In subsequent sections, we characterize several factors that lead some students to more quickly navigate the group search process in physics.  In particular, we reveal several major differences between students regarding level of interest development, when they started evaluating individual research groups, and the kinds of difficulties they experienced during their group search process.

\begin{figure*}[t]
\centering
\includegraphics{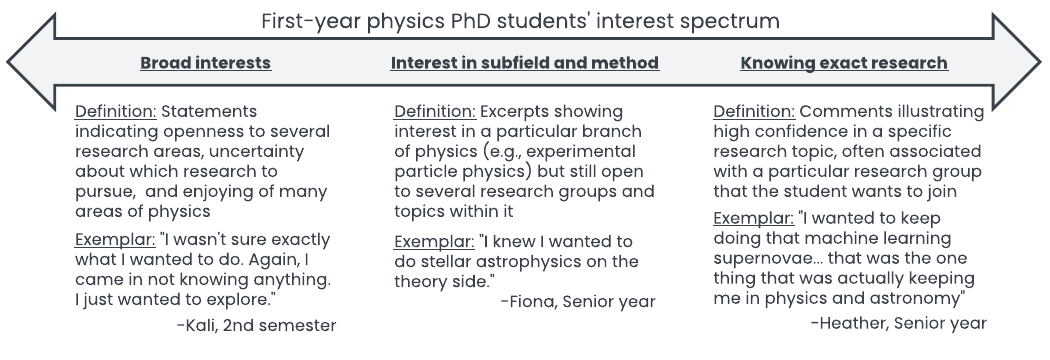}
\caption{\label{fig:interest_spectrum} A summary of codes used in Results Section \ref{subsec:interests}.  We conceptualize physics PhD students' certainty regarding their research interests as existing on a spectrum.  For this analysis, we grouped students into three categories: broad interests, interest in subfield and method, and knowing exact research.  Students with broad interests were less sure of what research they wanted to pursue in graduate school, while students who knew the exact research they wanted often already had a research group in mind upon entering graduate school.}
\end{figure*}

\subsection{\label{subsec:interests} Some physics PhD students know what research they want to do, but many do not}

One broad theme that emerged from the data was that graduate students' certainty about their research interests varied widely, and that students' interests were often still malleable when they began graduate school.  We identified three sub-themes within the broad \textit{Interest development} code: \textit{Broad interests}, \textit{Interest in subfield and method}, and \textit{Knowing exact research}, represented as a spectrum of interest development illustrated in Figure \ref{fig:interest_spectrum}.

Since students' experiences searching for a research group often spanned several semesters, some interviewees described how their interests shifted over time.  Some students made statements coded as \textit{Broad interests} during their junior year of undergrad, but later described how other factors (e.g., research experiences, talking with advisors) caused them to become interested in particular topics.  These students have excerpts appear in multiple categories of interest, but at different times in their narrative.  Thus, the sum of students who made comments coded under each of the categories is higher than the total number of students in the sample.  
% Figure \ref{fig:venn} breaks down each category to elucidate how many students were uniquely coded under a particular sub-code, or whether they appeared under multiple categories.  For example, the figure illustrates that 15 students \textit{only} had excerpts coded as \textit{Broad interests}, indicating that they had entered graduate school open to a variety of research topics.  Meanwhile, a different 6 students had excerpts in both the \textit{Broad interest} and \textit{Interest in subfield and method} categories, indicating that these students became more confident in their interests over time.   

%\subsubsection{Broad interests}

\textbf{Broad interests} Many students ($N=24$) at some point in their narratives said that they were open to pursuing a variety of research areas and had a broad set of interests in physics.  These included statements in which students expressed uncertainty about what research they wanted to pursue, as well as comments about how they believed they would enjoy several subfields and methods.  For instance, Nathan described how ``It's very easy to get me interested in things, especially in physics. So the whole time, my thought has been to keep an open mind.''  Two students, Valerie and Kali, explicitly said that their goal in graduate school was to figure out what they wanted to research.  Kali stated that ``I just wanted to explore,'' while Valerie said that ``I wanted to go to grad school because I didn't know what I wanted to do in physics. And I wanted to figure it out in grad school.''  Thus, both entered graduate school with the expectation of being able to explore several interest areas.  

Other students described more specific indecision, such as whether they wanted to pursue more or less hands-on work. Hassan recalled that coming into graduate school, ``I was kind of not really sure what I wanted to do, theory or experimental research.''  Ash had nearly the same sentiment, recalling that ``Coming in, I was deciding if I wanted to do high energy theory or high energy experiment.''  Meanwhile, Cole described having interests in quantum-related topics, but did not have a strong preference for the kind of research he wanted. He described searching for graduate schools based on the idea that ``I wanted to do things with quantum stuff, maybe use machine learning or something, do something wacky.''  However, as Cole later noted, many subfields of physics involve ``quantum stuff'' and use computational techniques such as machine learning, which left numerous research pathways open to him.

A significant subset of students who reported broad interests ($N=8$) did so during or after their first semester in the graduate program. Kali recalled that during her second semester of graduate school she still ``wasn't sure exactly what I wanted to do... again, I came in not knowing anything.''  Greg remarked that by his second semester, he was still ``open to doing whatever research in grad school, whether it be biophysics or condensed matter. I wasn't like, heart set on one.''  Meanwhile, Isaac noted that in his second semester, ``I didn't really know what I was interested in... all I knew was, like, something that involved quantum to some extent.''  Thus, these students entered graduate school without a clear idea of which research they wanted to pursue and had not clarified their decision during their first semester.  Most often, these students were too busy with other responsibilities to worry about finding a group.  Per Isaac, ``First semester, finding an advisor was kind of like the last thing on my mind... [friends] would ask me sometimes, have you found a research group? And I'd be like no, I haven't even thought about it. I'm trying to figure out the electric field of a charged hemisphere, or whatever the homework was about.''        

% \begin{figure}[t]
% \centering
% \includegraphics[width=\columnwidth]{venn.pdf}
% \caption{\label{fig:venn} Diagram illustrating the number of unique students who had excerpts coded under each interest code. Since students' narratives spanned multiple semesters, they often expressed how their interests developed over the course of their timeline.  For example, the 6 in the diagram indicates that six students had described having broad interests at one time, but later had become more interested in a particular subfield and method.  Notably, 15 students only ever discussed having broad interests.  Numbers represent unique students, so their sum is 40.}
% \end{figure}

%\subsubsection{Interest in subfield and method}

\textbf{Interest in subfield and methods} It was also common for students to indicate interest in both a subfield and method while narrowing their search for a research group ($N=23$).  Excerpts coded as \textit{Interest in subfield and method} indicated a stronger affinity for a particular area of research than excerpts coded under \textit{Broad interests}.  For example, Elena expressed that because she had already worked in high energy experiment in her undergraduate research, ``I was pretty secure in the fact that like, this is really what I wanted to do. I work in experimental high energy physics. And so I felt like I had kind of that track already laid out.''   Likewise, while applying to schools his Senior year, Alex affirmed ``My field is experimental optics, so I generally knew that I wanted to do something with laser sources, and applications of laser sources.''  Nina also described hoping to stay in the same research area (high energy experiment) for her PhD, saying that ``When I was trying to apply to the US, I was sort of trying to stick to the, either the same experiment that I've been working on, or something similar.''

% Despite expressing clearer interests, several students were still unsure of which exact topic to pursue within those areas.  Although Luis was confident that he wanted to pursue experimental high energy research, he recalled weighing several research topics under that umbrella: ``That was around the time when I decided, neutrinos are cool, but my interest at the time was really accelerator physics. Like doing stuff with the LHC.''   Similarly, Gabriela described her ``only requirement'' for research ``was that it was hands on experimental condensed matter.''  Although this helped Gabriela narrow her search, she was ``mainly just looking for the right PI'' and was therefore willing to be flexible with her particular research topic.

%\subsubsection{Knowing exact research}
\textbf{Knowing exact research} A smaller subset of graduate students in our sample ($N=13$) reported being highly confident in the research that they wanted to pursue.  For instance, Heather said that being able to work on machine learning and supernovae was ``the one thing that was actually keeping me in physics and astronomy.''  Indeed, upon entering graduate school and beginning to work in her current research group, she affirmed,  ``This is exactly what I want to be doing in astronomy. Like, if I decide this specific thing isn't for me, there's not really anything else I'd rather be doing here, y'know?''  Hence, she suspects that if things ``didn't work out'' with her current advisor, ``I probably wouldn't stay in grad school. Like, at that point, I would just go into industry.''  Heather's comments highlight how strongly some students' research interests can influence their willingness to persevere in graduate school.  

Most commonly, students described a strong interest in working with a particular faculty member at their prospective graduate school.  For instance, Rose described already having a ``person of interest'' with whom she wanted to work in graduate school, while Lakshmi said she ``had it pretty sorted in my mind where I wanted to go'' and who she wanted to work with before arriving.  Similarly, Selena recalled that ``I'm pretty sure my whole admissions essay was just, I want to work with [this professor]. And also, these other people are cool, but mostly [him]. Yeah, I sort of came in very much angling for that.''  She noted that this certainty was based primarily on this professor's research topic, saying ``I didn't really know much about him other than just what he did. You know, like what's available on a website for a professor. But it definitely was like, this is stuff I think is really cool.''  

While Selena's interest in this faculty member was driven primarily by her perceived interest alignment, others attributed their confidence in wanting to work with a particular faculty to other reasons as well.  Ursula had worked with a particular faculty member through a collaboration in her undergraduate research, which gave her confidence in her ability to work in his group.  ``The head of [the collaboration's] gravitational waves group is actually the person that I applied to work with here.''  She continued, explaining how her prior experience informed her that he would be a good mentor: ``I knew that if I were to work with him, I'd have to kind of relearn a lot of things... but it felt like with his guidance, and with his wealth of knowledge, and willingness to explain things, and excitement about explaining things, it felt really worth it.''

\begin{figure*}[t]
\centering
\includegraphics[]{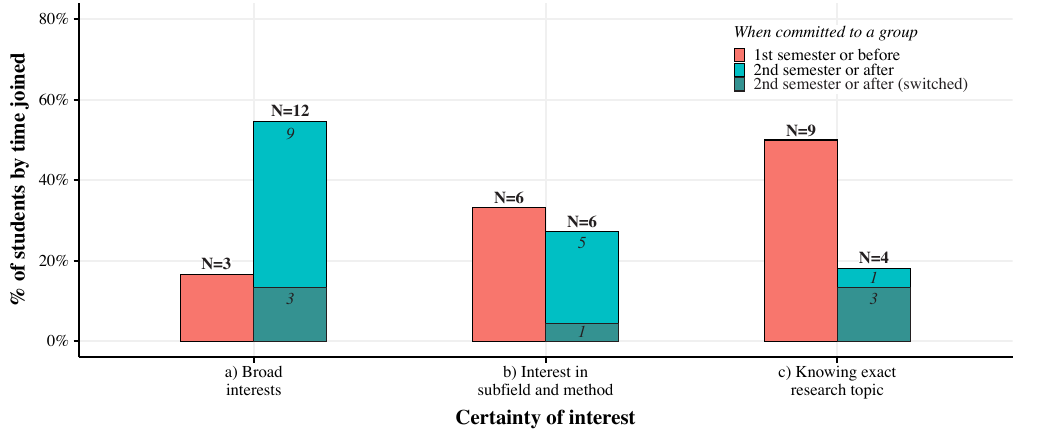}
\caption{\label{fig:interests} Students with broader interests tended to join research groups later than students with narrower interests.  The x-axis breaks down students by the highest level of interest they described upon entering graduate school. Bars represent the percentage of students in each interest category who joined 1st semester or before (red) or 2nd semester or after (green).  Thus, the sum of the red bars and green bars add up to 100\%, respectively.  The dark green indicates students who had joined a group early in their graduate careers but switched later on.}
\end{figure*}

\subsection{\label{subsec:intereststime}Students with narrower interests join research groups more quickly}

Having established a spectrum students' research interest certainty, this section examines differences in when students with varying levels of interest joined their research groups. To do this, we sorted students into groups based on the highest level of interest they expressed before starting graduate school ($N=15$ in \textit{Broad interests}, $N=12$ in \textit{Interest in subfield and method}, and $N=13$ in \textit{Knowing exact research}).  We then split each interest group by when they committed to a research group and compared student counts. 

As illustrated in Figure \ref{fig:interests}, we observe that students in our sample who expressed having broad research interests upon entering graduate school were more likely to join a group in their second semester of graduate school or later.  Figure \ref{fig:interests}a) shows that 12 of 15 students who were categorized as having broad interests found their research groups during the second semester or later, compared to 3 of 15 who joined in their first semester or before.  The green bar of \ref{fig:interests}a) also demonstrates that of all 22 students who joined a research group during or after their second semester, 12 of them (54\%) had been coded as having broad research interests.  This is compared to the red bar, which illustrates that of all 18 students who committed to a group first semester or before, only 3 (17\%) were coded as having broad interests.  

Comparing Figure \ref{fig:interests}a) to Figs. \ref{fig:interests}b) and \ref{fig:interests}c), we see that the time to join a research group decreases as certainty of interest increases.  Figure \ref{fig:interests}b) shows 33\% ($N=6$) of students who joined first semester or before expressed interest in subfield and method, compared to 27\% ($N=6$) who joined after first semester.  Meanwhile, in Figure \ref{fig:interests}c), we see that students who came into graduate school with strong research interests tended to commit to a research group early.  Indeed, 50\% ($N=9$) of all first semester or earlier joiners were students who described knowing the exact research topic they wanted to pursue.  This is compared to just 18\% ($N=4$) of the students who found a group during their second semester or later.

In addition to sorting students by when they joined a research group, Figure \ref{fig:interests} also indicates whether the student had switched out of a different research group prior to joining their current lab.  Of the four students in Figure \ref{fig:interests}c) who came in with strong interests but joined later, three (Benjamin, Kwan, and Tabitha) had switched research groups.  This shows how these students had indeed joined a research group quickly upon entering graduate school, but discovered that their groups did not fit them well.  For example, Benjamin began working with a research group his senior summer in order to pursue his interest in fusion research, but quickly realized that he did not like his work or the group's overall culture as much as he thought he would.  He recalled,  ``I don't know. I just wasn't super excited about the project... and there was like, Zoom meetings with a lot of cranky old scientists that were like yelling matches.  It was kind of a nasty environment.''  Thus, neither the research topic or group environment fit Benjamin's expectations.  However, since Benjamin had this experience so early in his graduate career, he was able to switch groups during his 2nd semester into one that better suited his interests.  Tabitha and Kwan's narratives are discussed at length in Section \ref{subsec:concerns}, where we more deeply examine the difficulties and concerns that students had while searching for a group.

On the other end of the spectrum, we see that three students who came into graduate school open to a variety of research topics also switched groups.  Two of these students (Cole and Eric) started to do research immediately upon arriving at graduate school, but did so with the goal of exploring their options for research and beginning to meet other graduate students. As described by Eric, ``I spent some time in [a professor's] lab, just a few weeks, helping out with stuff around the lab, just to, I guess, like, meet people and and get a sense of what the environment was like.''  Although Eric acknowledged that he might enjoy the work and stick with the lab, he clearly anticipated the possibility that he would want to leave.  He recalled, ``I was definitely not confident that it would work out, which is part of why I was, very early, trying to see if I could hang out with [that] group for a while beforehand.'' For Eric, his first group amounted to a pseudo rotation period that he used to get a better sense of what he might want to do in graduate school.

\subsubsection{\label{subsubsec:ugresearch}Access to undergraduate research clarifies research interests} 

Aligning with prior literature discussed in Section \ref{sec:Introduction} on the impact of undergraduate research, students in our sample with stronger interests often cited their undergraduate research experiences as a major factor impacting their interest development.  For example, Elena's comment that working in high energy experiment as an undergraduate made her feel that her path had ``already been laid out.''  Similarly, Desiree described her undergraduate research working in high energy theory as the ``key event'' in her interest development, saying ``That was when I was like, okay, this seems like what I want to do.'' Fiona noted that she ``did some research where I did my bachelor's, and then I also did the two REUs. And from that I knew what I was interested in was mainly the theory side of stuff, and specifically stars.''  All of these quotes illustrate how opportunities to engage with modern research areas in physics as an undergraduate make paths into those areas seem more open, and simultaneously open doors to pursuing them in the future. 

However, several students in our data cited struggles with finding a path into undergraduate research, which they associated with factors such as their demographic backgrounds or the small size of their undergraduate institution.  These findings are also in alignment with prior literature regarding access to undergraduate research experiences that were discussed in Section \ref{sec:Introduction}.  For example, Hassan was interested in topics related to quantum computing, but noted that his small undergraduate institution did not offer enough research opportunities for him to determine the kind of research he wanted to do.  He said, ``I didn't really have a lot of research background. So like, in undergrad, I got involved with like one project, which was small scale given the school that I attended. But I was not really sure what I wanted to do.''

One first-generation student in our sample, Carmen, relayed how ``In undergrad, I mean, I am a first-generation student. And I, as many people, struggled with my course load... I struggled to find someone who could advise me in research, and I applied to REU programs over the summer. My own department and externally. The first time I did that, I did not receive any offers.''  The following year, Carmen earned a research position in their home department doing astrophysics, indicating that this experience was the reason they were able to pursue a graduate degree:  ``I probably wouldn't be here to talking with you today [if not for the REU].''  Still, Carmen felt as though ``there were people who had, in my opinion, more resources and more exposure to the field at an early age than I did. And so I still didn't know what I wanted to do.''  Carmen entered graduate school unsure of the research they wanted to pursue and needing to find a group during their first year of graduate study.  However, as detailed in our previous work \cite{verostek2023physics}, this opened the door to a number of new challenges as Carmen searched for a group while trying to adjust to their new environment and taking graduate-level courses; we return to these aspects of Carmen's story in more detail in Section \ref{subsec:concerns}. %Carmen's experience is a microcosm of the broader message of this section, that inequitable access to formative research experiences can have long-term effects on the first-year graduate experience. 

Our data also revealed how access to advanced coursework and electives was another factor that shaped students' interests, helping them narrow their research interests sooner.  Although Desiree's interest in high energy theory was confirmed during her research experience, it initially stemmed from taking advanced electives: ``During my junior and senior years, I started taking more electives... particle physics, cosmology, things that are not in like the main courseload...  [topics] you literally never hear about in undergrad, unless you decide to go take a class on it.''  Similarly, Matias cited the graduate-level condensed matter course he took as a senior as being a pivotal influence, as he strongly disliked the undergraduate version.  He recalled, ``That professor was teaching graduate solid state, and I was like oh shit, here we go again. But I enrolled. And then when we looked at it from a graduate perspective, I started to like it.  I was like, wait, okay, actually this is kind of cool. We looked into topics in a deeper sense you know?''  Yet students from smaller departments, particularly those without a graduate program, might not offer such courses at all.  Collectively, this section indicates some of the ways that a student's background and access to advanced coursework and undergraduate research opportunities might impact their search for a research group down the road as . 

%For instance, Hassan had described a lack of access to research at his undergraduate school that contributed to his uncertainty about what research he wanted to pursue in graduate school. 

% Despite the predominant trend that students who express a high degree of certainty in their interests tend to join groups more quickly, we see also observe several deviations from this tendency. Four students who described entered graduate school knowing the exact research they wanted to pursue ended up joining a group sometime after their second semester.  For one of these students, funding issues for her preferred advisor meant she had to wait until the summer of her first year to begin work with him.  The other three began working in their preferred research right away, but negative experiences drove them to switch groups.  These students' narratives are discussed at length in Section \ref{subsec:concerns}, where we more deeply examine the difficulties and concerns that students had while searching for a group. 

\subsection{\label{subsec:stages} Early connections with prospective advisors help students join groups sooner}

\begin{figure*}[t]
\centering
\includegraphics[]{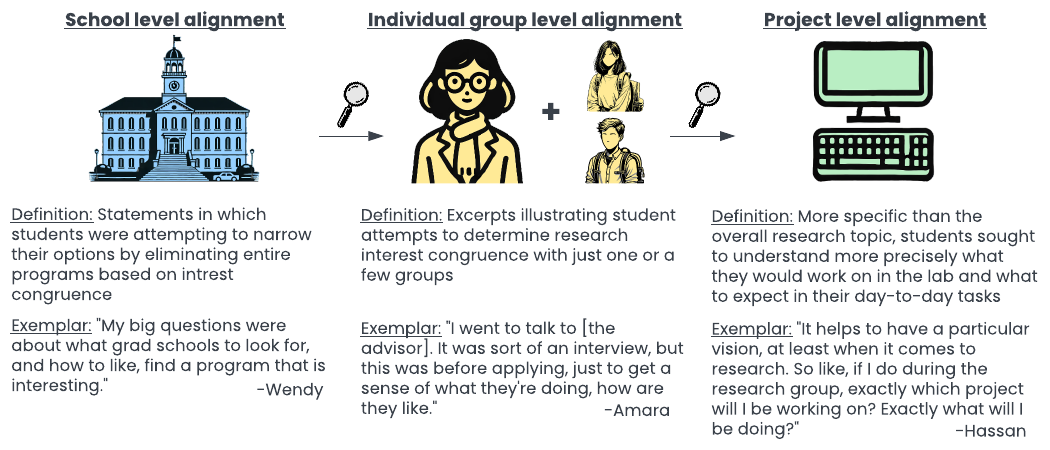}
\caption{\label{fig:stages_of_searching} A summary of codes used in Results Section \ref{subsec:stages}.  They illustrate three broad stages students described going through while trying to find a research group with congruent research interests.  Students generally described a narrowing of choices beginning with entire programs at the school level.  Once a few schools were identified, students could more carefully evaluate their alignment with individual research groups within those institutions.  A subset of students were specifically interested in the exact project that they would work on if they were to join.}
\end{figure*}

Another major factor that \citeauthor{lovitts2002leaving} cited as predicting success in the PhD was students' proactivity in searching for a research group \cite{lovitts2002leaving}.  Specifically, she found that in the period before arriving at graduate school, students who collected information about individual faculty advisors were more likely to persist.  Students who contacted faculty directly or went for on-campus visits were more likely to complete a PhD than students who relied more on the school's reputation or looking at faculty websites:  ``the closer the source of information to the most interpersonal interactional level (faculty), the larger and more significant the difference between completers and noncompleters becomes.''  In this section, we illustrate the variation in when students start earnestly looking into research groups, and show how students who engage in the search process earlier tend to join a group more quickly.     

For this analysis, we split students' stages of evaluating research alignment into two categories based on the specificity of their search. These were \textit{Evaluating school level alignment} and \textit{Evaluating individual group level alignment}.  We also generated a sub-code within the larger individual group alignment category called \textit{Evaluating project level alignment}.  As was the case in Section \ref{subsec:interests}, students frequently had excerpts appear in several of these categories.  For instance, a student may have started narrowing their search during their junior year by trying to figure out which schools have an astrophysics program (\textit{School level alignment}).  Then once they were accepted to several programs, they might have started to think more deeply about which particular groups interest them (\textit{Individual group level alignment}).

\textbf{School level alignment} Most students began narrowing their search for a group by trying to determine whether departments had several professors working in an area they might be interested in.  Excerpts coded as evaluating \textit{School level alignment} indicated that students were attempting to narrow down their options by eliminating entire programs, rather than trying to decide between a handful of professors.  Most commonly, students leveraged their experiences in undergraduate research and online resources such as school websites to help in this endeavor.  Wendy, a first-generation graduate student who majored in physics and computer science, explained that her search for a research group began with trying to find a program with computational astrophysics: ``My big questions were about what grad schools to look for, and how to like, find a program that is interesting.''  She also noted that ``I spent a lot of time figuring out, even the words `computational astrophysics,' I didn't know what that meant for a while. Like, I didn't know what the thing I was looking for was.''  However, once she determined that computational astrophysics was the topic she wanted to pursue in graduate school, she began ``Googling like crazy'' and eliminating programs that did not list computational astrophysics as an active research area.  

Students with broader interests than Wendy indicated that their criteria for narrowing down schools was not as stringent.  Isaac said that while applying to schools, ``I knew vaguely what area of physics I was interested in... I guess, at the time, particle physics, and maybe like quantum field theory, or something. So my thought was just, you know what, let's see if there's anyone out there doing that.''  Online resources also played a big role at this time for Isaac, who said he would ``go to [the physics department's] page, look at the professors, and just click on people who seemed interesting.''  A common refrain from students at this point in their process was that they wanted to identify programs with several groups whose research seemed to align with their interests.  For instance, Blake said that they were drawn to their graduate program because ``because there's a lot of research in what I wanted to do here,'' which was appealing because ``I didn't have a super specific focus I wanted within optics.''      

% Several students, like Matias, wanted to know that there were a specific number of groups who seemed appealing ``I had this rule. It's like a general rule, I'm sure you've heard about it. Like, you want three professors you wouldn't mind working with at each school.''  Some students were knowing that there was a variety of research opportunities available at a prospective graduate school, and assumed that they would be able to find an advisor once they arrived.  

Many students ($N=28$) had excerpts coded in the \textit{School level alignment} category, and students were equally distributed over time to join a group (72\% who joined 1st semester or before and 68\% who joined after first semester).  This is indicative of how regardless of whether students were sure of what research they wanted to pursue, they still went through this process.  However, this means 12 students did not discuss looking at their school's research at all.  While several students may have simply omitted this from their narratives, some students specifically noted that they did not filter schools based on research.  For instance, two students attended the same graduate school where they went to undergrad, and never contemplated going elsewhere.    Another student chose which graduate schools to apply for based on whether they offered a fee waiver, figuring they would sort out their research interests once they arrived.  Other students primarily filtered graduate schools based on their location and proximity to family.  Meanwhile, several students strongly coupled their choice of graduate school to an individual advisor, and therefore never talked about considering the research going on in the broader department.

\textbf{Indivdual group level alignment} Excerpts with student attempts to determine research interest congruence with just one or a few groups were coded as  \textit{Evaluating individual group level alignment}. While the school level alignment phase commonly had students looking at school websites for help, this phase was most commonly characterized by talking to prospective advisors and graduate students directly.  In both stages, students reflected often on their undergraduate research experiences for help deciding whether a research topic seemed interesting.  In sum, excerpts from $N=34$ unique students were coded under \textit{Evaluating individual group alignment}, again indicating that students went through this process regardless of interest level.  

Students often described becoming more focused on particular groups once they arrived in graduate school.  For instance, Valerie recalled beginning to ``go between research group to research group, all these different meetings... definitely comparing them against each other, and then comparing them against the good internship experience I had.''  Hassan said that it was ``first semester when I slowly started getting more into, like the in-depth portion of the research that [this faculty member] did.''  This involved meeting directly with his prospective advisor: ``I went and met with him. We briefly talked about what he does, I got a tour of the lab.''  Jiya emphasized that one of the primary benefits of being on campus with prospective advisors was that she could ask them directly about their research, saying ``I want to hear it from them, like how they describe it.''

However, many students emphasized that they had already made efforts to evaluate individual groups \textit{before} they arrived for their first semester of graduate school.  For some, this was an integral part of choosing which school to attend; as described by Chloe, ``for me, choosing a school was pretty much choosing a research group.''  In an effort to figure out who she wanted to work with before accepting an offer, she described ``the hell of meeting with every single faculty member and every single one of their grad students and every single one of their co-workers'' while finishing her Senior year of undergrad, which made her feel ``immensely anxious.'' These meetings with faculty were her main source of information regarding each group's research activities: ``For specifically doing the research I wanted to do, that primarily came from the advisors.''  Beyond the interpersonal connection offered by just contacting professors, Irene described the benefits of being able to try working in a lab the summer before graduate school.  She recalled how she was ``very, very stuck in a rut'' between wanting to work with two individual faculty members, but when one of them reached out to her and offered her a summer research position prior to starting graduate school, Irene took the opportunity.  ``At the end of the summer, [the professor] was like, you know, if you want to join our group, you can join our group.''       

Within the category of looking at individual groups, several students described specifically wanting to evaluate research at the \textit{Project level}.  Excerpts coded under this category show how students sought to understand more precisely what they would work on in the lab if they were to join, and what to expect in their day-to-day tasks.  For example, after being accepted to his top choice graduate school in his Senior year of undergrad, Alex recalled asking professors, ``Where do you see your research going in the next three to six years? Is the work that you're doing now going to continue, or am I going to join your group and find that you're pivoting 180 degrees? And you know, now you're not doing the work that I joined your group to do.''  Matias similarly wanted to know what the ``day-to-day'' of his work would be like in a prospective group, which he was able to answer by earning a fellowship to work there the summer before graduate school.  He said, ``I would say if I didn't have that opportunity, I would still be very uncertain right now.''  Comparatively fewer students ($N=15$) made statements coded under this sub-category, suggesting that it may not be a topic that students looking for groups consider as often or feel that they are able to evaluate before deciding to join a group.

\begin{figure}[t]
\centering
\includegraphics[]{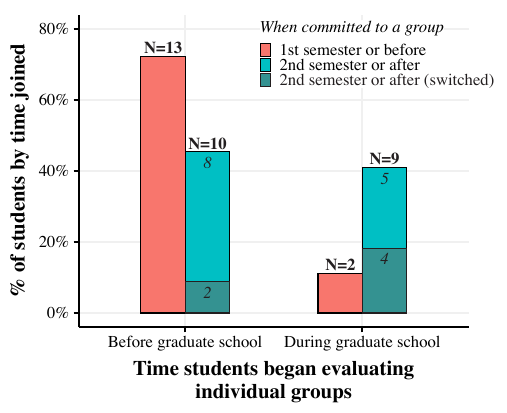}
\caption{\label{fig:stages_bytime} Students who engage in the search process earlier tend to join a group more quickly.  The x-axis splits students by whether they had begun evaluating individual research groups before or after starting graduate school.  Bars represent the percentage of students in each interest category who joined 1st semester or before (red) or 2nd semester or after (green).  Dark green indicates students who switched research groups.  In this case, the bars do not add to 100\% because not all students had excerpts coded under evaluating individual groups. }
\end{figure}

In total, $N=24$ students talked about evaluating their interest in individual groups before graduate school, while $N=11$ exclusively talked about looking at research groups after arriving in graduate school. Figure \ref{fig:stages_bytime} plots these two subsets of students, split by when they joined a research group. We see that students who joined a research group during or before their 1st semester were more likely to have evaluated individual groups or projects before they arrived in graduate school than students who found a group second semester or later.  72\% ($N=13$) of students who committed to a research group in their first semester or earlier had started the process of evaluating individual research groups before graduate school, compared to 45\% ($N=10$) of students who joined second semester or later.  Only 11\% ($N=2$) of the earlier joiners indicated that they had not already been looking at specific groups prior to arriving in graduate school.  On the other hand, 41\% ($N=9$) of those who joined second semester or later did not begin closely looking into individual groups until starting their programs.  This illustrates the variation in when students start earnestly looking into research groups, and shows how students who engage in the search process earlier tend to join a group more quickly.

Figure \ref{fig:stages_bytime} also highlights the subset of students in our sample who switched research groups, shown in dark green.  The two students represented by the dark green bar on the left are Benjamin and Tabitha, both of whom were also categorized as knowing the exact research they wanted to pursue.  Benjamin was guided toward his research group by his undergraduate advisor, who was a collaborator with his prospective graduate advisor; Tabitha's prospective advisor contacted her directly to recruit her to his group. However,  both students found that their graduate labs did not fit them well and switched.  

Meanwhile, Cole and Eric are among the four switchers represented by the dark green bar on the right of Figure \ref{fig:stages_bytime}.  As discussed in Section \ref{subsec:interests}, Eric and Cole came into graduate school with broad interests.  Yet despite having little contact with the professors beforehand, both started working in research groups shortly after arriving.  This illustrates a major difference from many other students in our sample, as Cole and Eric were willing to join a group quickly without much looking.  They intended to explore and likely switch groups later, which is an option that other interviewees with broad interests did not do (or did not know was an option at all).     

\subsubsection{\label{subsubsec:ugconnections}Connections from undergraduate experiences help students search earlier}

The resources that students used while looking at schools or contacting individual research groups varied.  Websites helped to figure out which schools might have interesting research, but students often had difficulty gleaning more specific details about a professor's research solely from online resources.  As noted by Irene, websites were ``the first point of information. Like when I was first doing research for schools, I would go to the school web page, go to the faculty and look at all their websites. And a lot of them if you go down to the bottom, it says last updated 2015. At that point, you can take none of the information.''  For more detailed information, other sources of help like as one-on-one conversations with faculty were significantly more valuable.  

At the same time, websites are significantly more accessible and less intimidating than meetings with graduate faculty.  Prior to arriving in graduate school, setting up meetings with prospective advisors with whom students are not co-located was more difficult and required more effort.  Matias discussed the lack of time associated with finishing undergraduate coursework at the same time as applying to graduate school: ``I barely had time because I'm finishing up my semester, I did a double major so every semester was like 20 units, so I was like, up to here with work. And so it was really, it's really hard to set time aside to look into these things, you know? Also... I was like, I'm not even gonna get in. So what's the point of looking into it?''  Some discussed anxiety associated with reaching out to faculty. For example, Jiya indicated that she struggled with being able to ``put myself out there'' and start talking to faculty.  Isaac recalled that ``I guess the biggest question I had was, if I were to summarize, how to find someone I'm interested in and then how to talk to them and have them not think I'm an idiot.''       

On the other hand, some students cited their ease of access to faculty at their prospective graduate schools as helpful in navigating these difficulties.  For instance, Felipe's undergraduate research advisor helped him make connections with his eventual group in graduate school: ``One of [my mentors] encouraged me to apply here. Because he knew the people in the research group I was interested in working with, so it was like, seems like a nice fit.''  Felipe leveraged his undergraduate advisor's connections to break the ice with his eventual graduate advisor.  Similarly, Alex said that emailing faculty was easy after his REU, which he described as ``a program whose sole purpose was to help students get into [the school, so I] had those connections essentially made for me. So when I wanted to take advantage, all I needed to do was send an email, basically no introduction necessary.''  The networking opportunities that undergraduate research provided allowed some students to more efficiently get in touch with individual groups.  Constastingly, Selena cited her background at a small school as a distinct disadvantage during her search. ``My school was small enough, we didn't have connections to these places I was applying. So it wasn't like I could say to the person I'm working with in my school, can you reach out to someone there? Can you help me find someone?  So it was very much like Google, and a lot of clicking, was how I tried to piece it together.'' 

These results suggest some of the resources used to more closely evaluate research groups (e.g., meetings with faculty) may not be as easily accessible to some students as others.  Some are less likely to have the time or connections to look beyond the school level while narrowing down groups,  Or, as detailed in our previous work, they simply may not know whether activities such as reaching out to faculty before starting their graduate program are appropriate \cite{verostek2023inequities, verostek2023physics}.  

% As illustrated in Figure \ref{fig:stages_bytime} however, students who are positioned to start looking at individual groups before graduate school are more likely to join a group sooner.  Moreover, since the number of research groups available for students to join in a department is limited, a student's access to their desired research group may end up depending on how quickly they make contact with a faculty member.  Indeed, Gabriela posited that this was the reason for getting into her group ahead of several other students  ``I think it's just because I reached out [to my advisor] early.''  This dynamic is explored more in the following section, which deals with several of the specific difficulties and concerns that students expressed during their search for a group.  

\subsection{\label{subsec:concerns} Difficulties and concerns in the group search}

Thus far, we have observed that students who joined research groups in their first semester or before tended to have narrower interests and began looking at groups sooner than students who joined second semester or later.  These results suggest that students' interest development and proactivity in the search process are important constructs in a model of how and why students join research groups.  Our results have also illustrated several benefits perceived by students who joined a group earlier in their graduate careers.  These included a sense of security knowing they did not have to search for a group during the first year of coursework, and in some cases an accelerated timeline toward graduation.  

Here, we more systematically examine some of the benefits and drawbacks associated with students' time to join a research group by leveraging the results of previous work, which described some of the themes surrounding students' difficulties searching for a research group (e.g., \textit{Coursework and research in tension}, \textit{Wanting more ways to meet grad students and groups}) \cite{verostek2023physics}.  We also included several new categories that our prior work did not discuss (e.g., \textit{Funding}, \textit{Career concerns}), and refined others based on the inclusion of more interviews in our analysis (e.g., \textit{Doubting ability to do research} was previously part of \textit{Feeling unprepared}).  

A summary of the ten common categories of difficulties that students encountered, including definitions and exemplar quotes, is shown in Table \ref{tab:difficulties}. For each difficulty, we examined whether it was more or less prevalent among students who joined before their first semester or after.  The table was organized into sections based on whether the concerns were more commonly found among students who joined in the first semester or earlier, second semester or later, or were approximately the same.  Sections are sorted in descending order by code occurrence.          

\begin{table*}[htp]
\centering
\def\arraystretch{1.1}%  1 is the default
\begin{tabularx}{\textwidth}{>{\hsize=1.6\hsize}Z
                              >{\hsize=1.6\hsize}Z
                              >{\hsize=.4\hsize}Y
                              >{\hsize=.4\hsize}Y}
                              
%\cline{1-4}

\multicolumn{1}{c}{\textbf{Code applied and short definition} ($N$)} & \multicolumn{1}{c}{\textbf{Example excerpt}} & \textbf{Joined earlier} & \textbf{Joined typical}  \\
\hline \hline

\rowcolor[gray]{.9}[\tabcolsep] 
\multicolumn{2}{c}{\textit{Concerns more common for students who joined a group 2nd semester or later}} & \textit{N (\%)} & \textit{N (\%)} \\
\hline

\textbf{\textit{Funding (23):}}  
Instances of students indicating they needed to find out if a group had funding, or were concerned about competition for spots due to lack of funding
& \textit{``My position is not so well right now... his funding has dried out a bit. There are a lot of students who want to do a PhD under him... there are about 12 students already.''}
& 8 (44\%) & \textbf{15 (68\%)} (5 switched)  \\ \hline

\textbf{\textit{Doubting ability to do research (22):}}  \linebreak 
Student concerns about their physics knowledge and research skills, impacting their ability to understand the research going on in a lab or make progress in research. Imposter syndrome is often mentioned.   
& \textit{``Just a difficulty, maybe it's more of like imposter syndrome. Like, let's say I do join the research group, am I going to do well? Am I going to be smart enough to handle all these smart and complex topics?''} 
& 8 (44\%) & \textbf{14 (64\%)} (5 switched)  \\ \hline

\textbf{\textit{Coursework and research in tension (18):}} \linebreak  
Student difficulties with discerning how much time to dedicate to coursework or research/searching for a research group.  Included comments indicating that the content covered in classes is misaligned with research.
& \textit{``As far as thoughts and feelings about the process as I was working on finding something, I feel like a lot of that got swamped out by the other things that I was doing that first year. Just being very busy with classes and teaching.''} 
& 6 (33\%) & \textbf{12 (55\%)} (3 switched)  \\ \hline

\textbf{\textit{Difficulty communicating with advisors (16):}}   
Struggling to get in touch with advisors, often due to students' own anxiety spurred by desire to impress faculty.
& \textit{``I was concerned about making sure I asked the right questions about the papers and like, looked smart... being very scared of looking stupid in front of a professor.''}
& 6 (33\%) & \textbf{10 (45\%)} (3 switched)  \\ \hline

\textbf{\textit{Considering leaving the program (9):}} \linebreak 
Comments in which students recall wanting to leave their graduate program entirely.  Also includes discussion of topics adjacent to leaving, such as experiencing depression and doubting one's ability to complete the PhD.
& \textit{``That's also when I expressed to [my advisor] that I might not stay for the whole program, I might leave after a master's.'' }
&  2 (11\%) & \textbf{7 (32\%)} \newline (3 switched) \\ \hline

\textbf{\textit{Wanting more ways to meet grad students and groups (8):}} 
Descriptions of departments inadequately promoting communication between students and research groups.  Includes wishing there was a trial period for trying a group.
& \textit{``In theory at least, it seems like having some structure in place to be able to, if not work with groups, at least visit them in person and see what it looks like to work in that group would be helpful.''}
& 1 (6\%) & \textbf{7 (32\%)} \newline (2 switched)  \\ \hline

% \textit{\textbf{Career concerns (15):}}  
% Comments regarding how choice of research will support or hinder career goals (e.g., going into industry), awareness of certain career options
% & \textit{``So like, one of the most important things that I think about is, I mean, do people who do graduate school get permanent jobs?''}
% & 7 (39\%) & 8 (36\%)  \\ \hline 

\rowcolor[gray]{.9}[\tabcolsep] 
\multicolumn{2}{c}{\textit{Concerns with similar prevalence regardless of time to join}} & \textit{N (\%)} & \textit{N (\%)} \\
\hline

\textbf{\textit{Considering switching or switched groups (19):}}
Statements from students who are deciding whether to stay in a group or leave.  Includes concerns about leaving.  Distinct from \textit{Commitment to research}, which refers to comments before students started work in the group. 
& \textit{``So that was probably the biggest challenge for me, even trying to make up my mind if I want to stick to this group that I joined at the very beginning.''} 
& 8 (44\%) & 11 (50\%) \newline (7 switched)  \\ \hline

\textbf{\textit{Wanting more guidance on steps to find advisor (12):}} 
Remarks from students wishing that there had been additional resources from the department with expectations and advice on finding a research group
& \textit{``I think the difficulty was not knowing where to reach out for help initially...I just feel like that's part of the degree, right? Finding someone. So I feel like it should be should be readily available somewhere.''}
& 5 (28\%) & 7 (32\%) \newline (3 switched)  \\ \hline

\textbf{\textit{Hard time adjusting to graduate school (9):}} \linebreak 
Concerns about the structure of graduate school and feeling comfortable in the new environment overall, not necessarily related to the group search.
& \textit{``I didn't know a whole lot about grad school and academia in general... like, okay, I just submitted all my applications. Let me find out a bit more about what I'm getting into.''}
& 4 (22\%) & 5 (23\%) \newline (1 switched)  \\ \hline

\rowcolor[gray]{.9}[\tabcolsep] 
\multicolumn{2}{c}{\textit{Concerns more common for students who joined a group 1st semester or earlier}} & \textit{N (\%)} & \textit{N (\%)} \\
\hline

\textbf{\textit{Commitment to research (20):}} \linebreak  
Worrying about the long-term nature of research, particularly apprehension about devotion a specific research topic area 
&  \textit{``At this point it's like I'm committing to something for like the rest of my life. Oh my gosh, that's terrifying.''}
& \textbf{12 (67\%)} & 8 (36\%) \newline (1 switched)  \\ \hline

\hline

\end{tabularx}
\caption{\label{tab:difficulties} Summary of codes capturing the difficulties and concerns that students encountered during their group search.  Code definitions and exemplar excerpts are provided, along with a count of the unique number of students who reported each difficulty.  Counts are broken down by the students who joined a research group their first semester (Joined earlier) or before versus their second semester or after (Joined typical). Percentages represent the fraction of students within each group (e.g., 8 of 18 = 44\% early joiners reported funding concerns). The number of students who switched groups is also indicated.}
\end{table*}

\textbf{Differences across time to join} Many students ($N=23$) across our sample expressed concerns over \textit{Funding}, but they were more commonly voiced by students who found their group in the second semester or later.  Excerpts coded as \textit{Funding} concerns revolved around uncertainty in whether a group had funding, or students' perception that competition for a spot in a group was high due to limited funds.  For instance, Nathan recounted a meeting with a prospective advisor who did not have a spot in his group because another student had already approached the professor: ``He had one slot that was likely being taken by an incoming student who already had a master's. How am I going to compete with that?''  Several students suspected that competition for spots stemmed from admissions decisions.  Jack said, ``They have a very bad history of that... there's maybe four or five professors, and they all have maybe three students across all five years of the PhD, right? So it's maybe 15 spots overall. Some years they take 15 people in one year... And then people have to go and, they go do different areas of physics.''  This result aligns with earlier comments from students who reached out to research groups early and cited it as the reason they were able to get into their group ahead of others.

Another highly cited concern for many students ($N=22$) was \textit{Doubting ability to do research}.  These students described feeling unprepared for graduate research, and often referred to a sense of imposter syndrome, which is a common phenomenon amongst PhD students \cite{harvey1985if, holden2024imposter}.  As Wendy put it, ``Just a difficulty, maybe it's more of like imposter syndrome. Like, let's say I do join the research group, am I going to do well? Am I going to be smart enough to handle all these smart and complex topics?''  Meanwhile, when Pauline changed to a new research group at the end of her second semester, she said, ``I can't say I felt as confident because it was all, it's a lot of in-lab work,'' which she had never done before.  This code also included instances of students discussing difficulties understanding what a group's research entailed, which we detailed in our previous work \cite{verostek2023physics}.

These concerns were less common among students who joined a group earlier, who were more likely to have higher confidence in the research they wanted to pursue.  One major contributor to students' self-efficacy is experience successfully performing a task (e.g., a student who does research in an optics lab tends to feel more capable of doing that research in the future) \cite{bandura1982self}.   Students who enter graduate school having already done a significant amount research of research are more likely to have more strongly developed interests, and are also more likely to feel confident in their ability to succeed in future research.

Some students ($N=16$) also described \textit{Difficulty communicating with advisors} as an issue they experienced during their search for a group.  This difficulty predominantly revolved around students' own anxiety being a barrier to reaching out to prospective advisors.  Eric noted that worrying about impressing professors ``wasn't a huge barrier'' but ``definitely made me hesitate in sending emails.''  Meanwhile, Dev lamented that the process of contacting advisors ``feels asymmetric'' since ``we have to put so much effort to figure out what they're doing, while some of them don't even update their websites... there's no way to communicate except for them feeling like replying to our emails.''  Students who committed to a group first semester or earlier encountered this difficulty at a slightly lower rate.  However, none of these students shared the concern \textit{Wanting more ways to meet grad students and groups}, whereas six of the students who joined second semester or later were found under both codes.  This may suggest that the subset of students who joined later in the first year were more likely to put off reaching out, and therefore tended to want departmental support with that process.          

Indeed, we see that students who joined second semester or later were far more likely to discuss \textit{Wanting more ways to meet grad students and groups}.  Students who sought more structure meeting groups often talked about wanting ways to try different research, which aligns with the finding that these students also tended to enter graduate school with broader research interests and were less likely to have already connected with individual research groups.  Additionally, we see that this subset of students also tended to have more struggles with \textit{Coursework and research in tension}, as classes prevented them from dedicating time to their group search.  This lack of time may have played a role in wanting more structured ways to meet research groups, as their schedules prevented them from doing so on their own time.  Students who joined early occasionally cited this difficulty, but it was typically in regard to their desire to begin doing research in their lab rather than taking classes that they felt were not beneficial.    

Although many of the difficulties more commonly affected students who had not joined a group by their first semester, our results also reveal several drawbacks of joining a research group so early (the bottom of Table \ref{tab:difficulties}).  Most notably, there was a higher prevalence of students being worried about \textit{Commitment to research} before joining a group.  Olivia, who planned to begin doing research with a professor as soon as she arrived in graduate school, feared that she could ``get there and start talking about the research and doing it, and I might not like it at all... I don't really know the solution for that.''  Oftentimes students who had committed early to a group worried that they may be missing out on a better opportunity.  This even included students who expressed high confidence in their choice of research.  Ursula recalled that she ``was trying not to pigeonhole myself too early still, even though I knew I loved this subfield.''  In the end though, she committed to working with a professor whom she collaborated with during her undergraduate research.  

Lastly, we observe that students discussed having a \textit{Hard time adjusting to graduate school} and \textit{Wanting more guidance on steps to find an advisor} at similar rates, regardless of when they joined.  This result is consistent with our prior work indicating that regardless of prior experiences, students tended to recognize that there was little structure in place to guide students into research groups.  For instance, Elena joined her research group in her first semester, but noted how her ``network'' that she had built during undergraduate research ``really helped me not feel so alone, and not feel so like, who would I even talk to in the first place?''  It is also notable that among the nine students who described struggling to adjust to graduate school, four identified as first-generation students.  Among these four, two had considered leaving their programs and struggled to find a group.  The other two had the opportunity to participate in the McNair Scholars program, a post-baccalaureate program meant to increase the number of PhDs earned by underrepresented students.  They cited this as a major source of help in dealing with their uncertainties about graduate school and coaching them on when to apply and start looking for research groups. This illustrates how giving structure and guidance can have a major impact on students' graduate school success.

\subsection{\label{subsec:leavingsubset}Looking at the subset thinking about leaving}

As outlined in the introduction, attrition is a persistent issue in physics graduate education. Of the 40 students in our sample, $N=9$ identified a time during their narratives when they \textit{Considered leaving the program}.  Thus, their cases are particularly important to understand in order to better identify students most at risk for leaving.  As shown in Table \ref{tab:difficulties}, we observe that 2 of 9 students who considered leaving their programs were students who joined their research group in their first semester or before, suggesting that students who are able to begin research quickly were less likely to discuss leaving their programs.  This aligns with one of the major findings of our previous work, that students who struggled to navigate the search process tended to experience lower sense of belonging, and were therefore more likely to leave.

Among the subset of 7 students who joined after their first semester and who considered leaving, we now focus on two that highlight both the negative impact of struggling to find a group and the benefits that students can reap from a positive research environment.  In the context of results from previous sections, their narratives exemplify how students' background experiences can impact their search for a research group, and subsequently their decision to stay or leave their programs.  

Brianna and Carmen were both first-generation students who entered graduate school unsure of the exact research they wanted to pursue, a fact Carmen directly attributed to previous difficulties finding an undergraduate research advisor (see Section \ref{subsec:interests}).  As with many students with broad interests, Brianna had predominantly been evaluating research at the school level: ``I chose [the school] specifically because, not because of the research, but because of the proximity... and, I saw that they did have a good like three or four astrophysicists here.''  On the other hand, Carmen ``applied to a bunch of programs and got rejected from all of them,'' but was able to earn admittance to an institution through the APS Bridge program.  With regard to finding a research group though, Carmen saw their position as disadvantageous: ``I was excited to have been admitted, but it's not the traditional route where you know who you're going to work with. That's a common story that I hear, right? But I did not know who I was going to work with.  I was just taking it one moment at a time, because that's the cards that life had dealt me.'' 

Upon entering graduate school, Brianna and Carmen grappled with many of the challenges enumerated in Table \ref{tab:difficulties}.  Seeing other students who had already joined groups drove concern that they were already falling behind.  As Brianna recalled, ``When I saw that everybody else had advisors, I was kind of freaking out, because I'm like, how? They just got here, how do they already have advisors?''  Seeing others seemingly ``succeed'' in the program where they had not contributed to lingering \textit{Doubts about ability to do research}.  Indeed, Carmen described entering graduate school feeling like ``I wasn't sure if I would be successful or not.''   Such concerns were exacerbated by a \textit{Hard time adjusting to graduate school}, which both Brianna and Carmen attributed to being first-gen students.  As Brianna said, ``I didn't know anything about grad school. I'm the first in my family to go to grad school. So I had no idea what I was doing.''  Lacking formal expectations, it was unclear whether she was on the right track or not.

As the first year progressed, both students experienced difficulties with \textit{Funding}, \textit{Wanting more ways to meet grad students and groups}, \textit{Coursework and research in tension}, and \textit{Difficulty communicating with advisors}, which contributed to their consideration to leave.  Being denied from groups due to funding constraints negatively impacted sense of belonging for both students. Carmen was rejected from the first group they approached: ``being my first attempt to reach out to a potential advisor, that was not a good feeling.'' Brianna went to four professors in her department who all ``said no, because they didn't have funding for another grad student... personally, I thought I was cursed.''  Carmen felt that the department did not do an adequate job facilitating interactions between graduate students, saying ``I also felt like I lacked a sense of community... yeah, a lot of my concerns were community based.''  Brianna recalled being confused as to ``whether or not I had to find [a group] right away, as opposed to like, waiting out and focusing on trying to pass the courses first.''  Opting to focus on coursework, she ``was under the impression that could wait till my second year,'' until the department informed her she needed a group by the summer, which made her feel ``taken aback. I was scared.''  Carmen described all of the first year responsibilities as ``more overwhelming than I had expected.''  Both described anxiety associated with seeking out help from faculty; as Brianna said, ``I just didn't know what was appropriate to ask and what not to ask... when you're talking to people who hold the fate of your career in their hands, it's like, I don't know if they're gonna think I'm dumb for asking this question. Because I've never experienced any of this before.'' 

In the end, both Carmen and Brianna were able to find research groups and remained in their programs.  Carmen was ``thankfully'' able to connect with an advisor through a voluntary departmental seminar series, but indicated that ``first year me'' could have easily missed it.  Brianna meanwhile described making a ``desperate plea'' to one more professor, who was able to fund her.  However, taken as a whole, these cases show how the difficulties most strongly associated with looking for a group while in graduate school can have major negative impacts on students.  Many students in our sample experienced these challenges, but for Brianna and Carmen, they accrued over the first year to an extent that pushed them to the brink of leaving. 

Our results have generally highlighted that that students who joined their research groups earlier in their graduate careers felt that it offered them a variety of benefits, such as avoiding competition for funding and alleviating anxiety in their first year.  However, a closer examination of the students who considered leaving their programs shows that this only tells part of the story, and illustrates the potential hazards of getting into a group too quickly.  

Leading up to graduate school, Tabitha and Kwan described experiences that were antipodal to those described by Brianna and Carmen.  Both Tabitha and Kwan came into graduate school knowing exactly the kind of research they wanted to pursue, and had been in contact with prospective advisors before arriving to their programs.  Kwan was guided to potential groups by his undergraduate research advisor, saying ``some other candidate professors made my list but I  screened them out by talking with my previous advisor. He was very supportive for applying to graduate school.''  Meanwhile, Tabitha's prospective advisor was a co-author on one of her undergraduate research papers: ``So one of the professors at [this school] reached out to me and he was like, `Hey, I read your thesis. We would like to keep on working on this and make it an actual paper' and I was like, sure.''  Thus, when it came time for her to find a PhD program, this faculty member was a natural fit.  She recalled him asking, ```Have you to considered going to [this school]?' Because he just happened to work here. Once again it's like, you want me, you want to have me as your student? I'm not gonna say no.''  

Thus, Tabitha and Kwan seemed to enter their graduate programs with several advantages relative to Brianna and Carmen.  Both expressed confidence in their ability to succeed in graduate school given their prior research experience.  Kwan noted that he had earned ``lots of achievements, I got a best paper award. And I got three journal publications under [my undergraduate advisor]... I presented a lot of international conference presentations and domestic conference presentations.''  They did not need to rely on their graduate departments to connect them with research groups, and they did not have funding issues.  Tabitha described her research as the ``perfect'' topic for her interests.  Yet in their second years as PhD students, both switched out of their research groups and were considering leaving their programs.

Tabitha's and Kwan's research topics matched their interests excellently, but their work environments drove them to leave.  From the outset Kwan felt that it was ``inappropriate'' for his advisor's significant other to be working in the lab, and felt that the advisor treated him unfairly because of this dynamic.  ``It was a kind of big problem, a big burden. And one day we just yelled at each other, this kind of serious argument.''  In his next meeting, his advisor said that their research style was ``not fit'' for him, and asked him to leave the lab.  Kwan, an international student, recalled, ``it was really stressful... I thought, why did I come to US?'' He continued, saying ``it was a kind very tough time for me. And I regretted about my decision to come to US a lot during my first year.''  Meanwhile, Tabitha became disillusioned with her group due to the poor working environment fostered by the advisor.  She said her advisor ``created a very toxic work environment... always trying to gaslight me into believing I was not good enough to be here... [trying] to force me to work outside of my working hours.''  Yet because she had come to the school specifically to work with this professor, she felt obligated to stay: ``It took me a while to realize that I was not in a healthy work environment.  Mostly because I was so, in a way, indebted to this professor for giving me the opportunity to come here. I think I thought, `Oh, yeah. He's not great, but I'm here because of him.''   More details of Tabitha's case are detailed in our prior work \cite{verostek2023physics}.  Ultimately, support from her cohort helped to convince her that she would be able to succeed if she stayed: ``I was doubting myself as a researcher. But my friends would tell me, no. It was you. You are doing the research.''  Although not the focus of this study, this quote suggests how a student's graduate cohort can also play an important role in supporting their sense of belonging.

Other students in our study had joined groups early and subsequently switched, but only Tabitha and Kwan indicated that they considered whether they would be able to continue in their programs.  Students who were motivated to switch groups in order to find research that better aligned with their interests did not indicate that their experience drove them to consider leaving their programs.  Indeed, for students who came in with broad interests and joined a group in order to try out the research, switching was an expected and positive move.  Consistent with prior literature on the impact of negative advising relationships, Tabitha's and Kwan's narratives demonstrate the outsized importance of group culture on student satisfaction.  They also illustrate the negative influence that joining a group so early can have on students' likelihood to stay in their programs if not properly supported.  Tabitha said of her experience, ``I definitely think that could have been avoided,'' and she wished that that someone in the department had told her earlier to rethink working for her advisor.  Reflecting on her search for a group, she emphasized to future students that ``It's a balance. It's hard to balance having a good advisor and having a research field that you're actually passionate about.''

\begin{figure*}[]
\centering
\includegraphics[width=\textwidth]{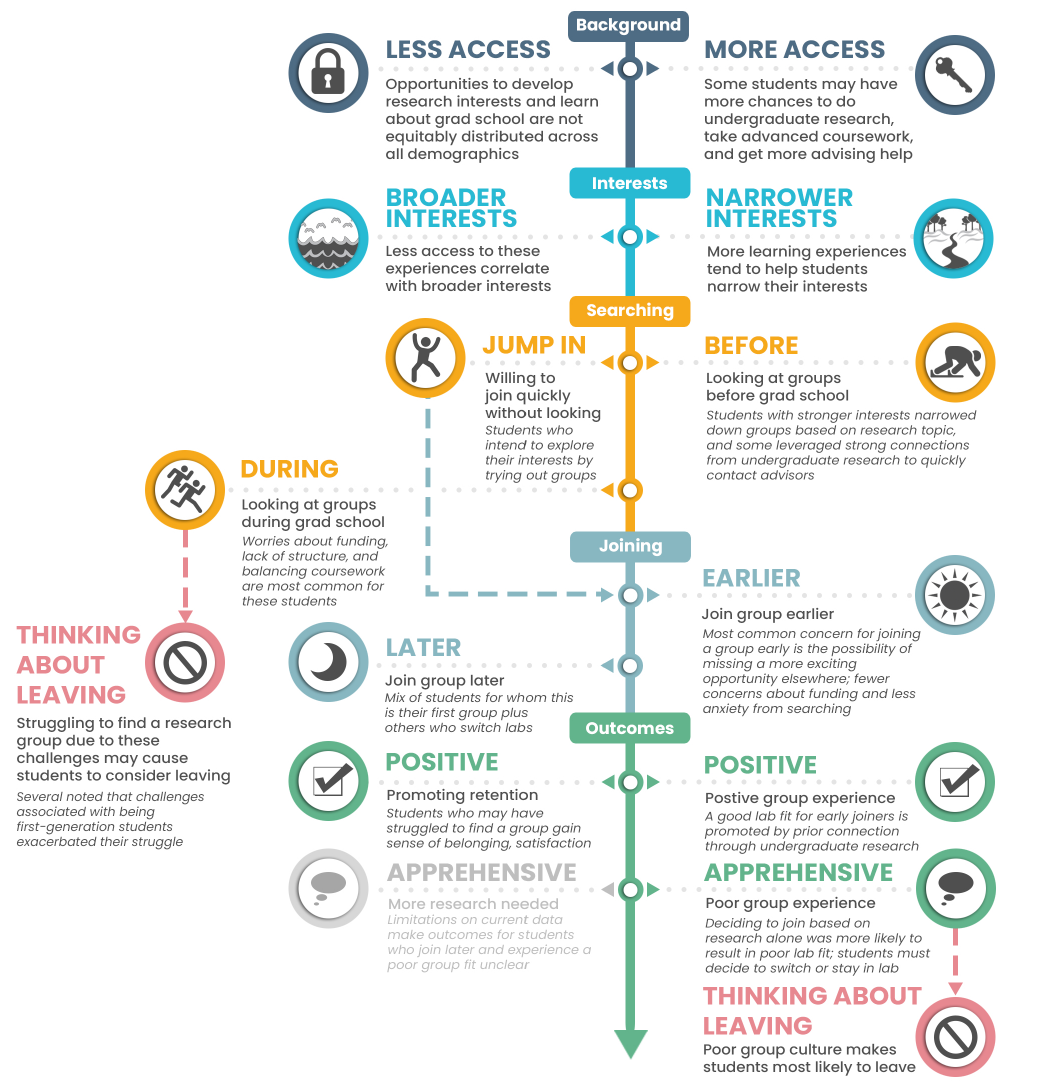}
\caption{\label{fig:discussion} A model of characteristic pathways that physics PhD students take during the process of evaluating the congruence of a group's research interests, exposing several underlying factors that contribute to students to thinking about leaving their programs.  This model illustrates how these factors interact and influence a range of student outcomes.  Student characteristics and outcomes on one side of the center dividing line are correlated with one other (e.g., broader interests are associated with searching during graduate school, which was correlated with thinking about leaving).  Despite the left-right divide, we conceive of each variable on in the figure as existing on a spectrum.  The grey text on the bottom left indicates an aspect of the model that we suspect exists, but was not present in our data.  See Figure \ref{fig:paths} for examples of how students in this study map onto the model.}
\end{figure*}

\section{\label{sec:Discussion}Discussion}

This study has examined the ways that physics PhD students' research interests influence their search for a research group in graduate school.  In this section, we synthesize our results in an explanatory model of student outcomes, shown in Figure \ref{fig:discussion}.  Grounded in data, the model highlights the connections between students' backgrounds, their research interests, when they search for a group, and the consequences of searching for a group before or during graduate school.  Moreover, by tracing students' journeys back to their undergraduate years, this model offers unique insight into some of the ways that inequitable access to formative experiences can affect outcomes down the road in graduate school.  Whereas prior research has correlated individual constructs (e.g., broad interests) with leaving \cite{lovitts2002leaving}, this model illustrates how these factors interact and influence a range of student outcomes.  Student characteristics and outcomes on one side of the center dividing line are correlated with one other (e.g., broader interests are associated with searching during graduate school, which was correlated with thinking about leaving).  Despite the left-right divide, we conceive of each variable on in the figure as existing on a spectrum.  For instance, some students may have some access to undergraduate research that helps them develop somewhat stronger interests, but not enough to tell exactly what research they want to pursue.

\begin{figure*}[]
\centering
\includegraphics[]{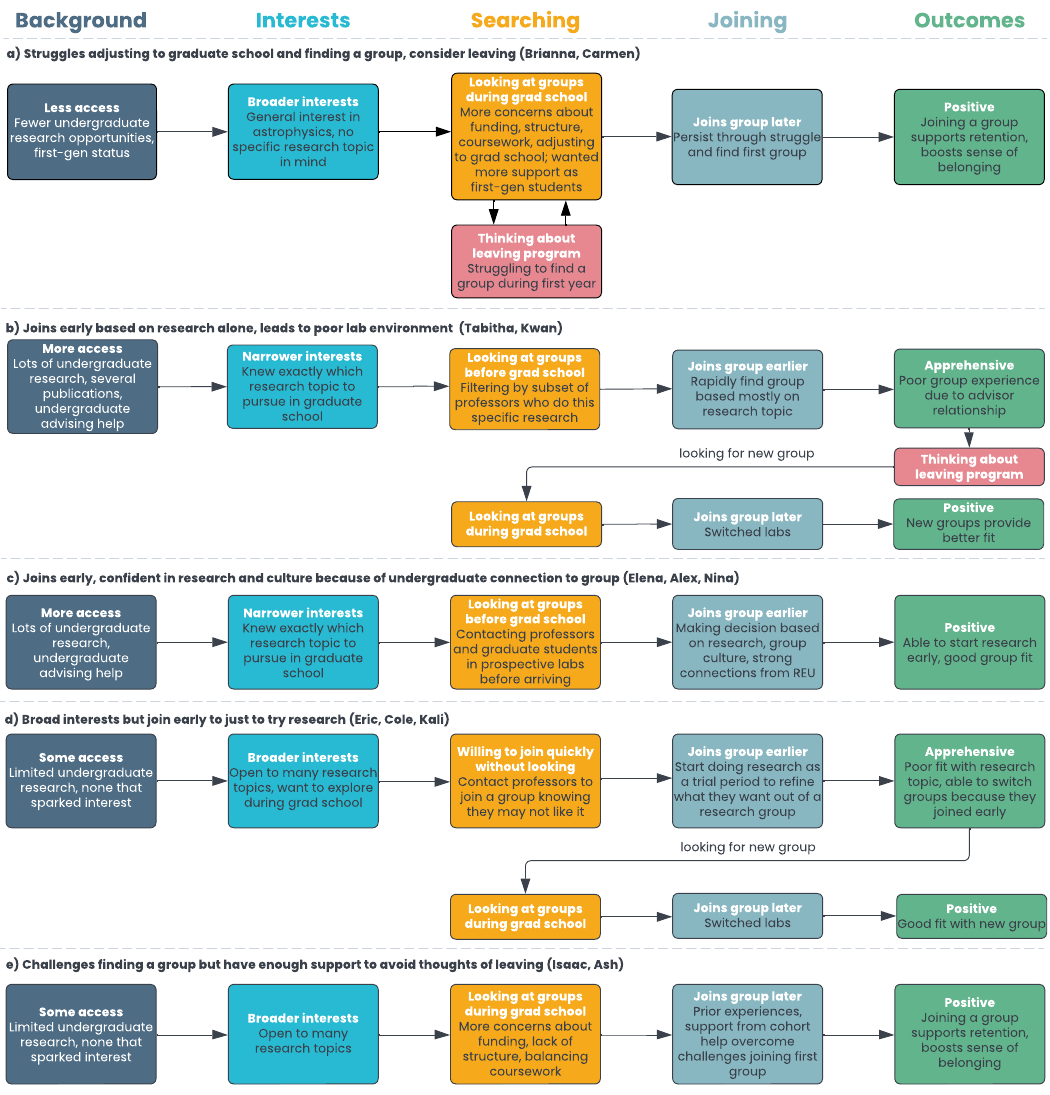}
\caption{\label{fig:paths} Several examples of students whose experiences searching for a group with congruent research interests that are well-described by the model in Figure \ref{fig:discussion}.  Each row represents a characteristic type of student experience, described alongside the student names in the study who fit that archetype.  The paths listed under a) and b) describe students who our results suggested were at a higher risk of leaving their programs.}
\end{figure*}

Figure \ref{fig:paths} provides several examples of how students exemplified throughout the study are indeed well-described by diagram shown in Figure \ref{fig:discussion}.  Critically, our work offers insight into two characteristic pathways that we observed as placing students at a higher likelihood of leaving their programs, highlighted in Figs.\ref{fig:paths} a) and b).  Figure \ref{fig:paths}a) represents the path exemplified by Brianna and Carmen. Both were first-generation students who entered graduate school unsure of what research they wanted to pursue.  Thus their group search took place in graduate school, where they encountered a number of difficulties that impacted their overall sense of belonging in the program.  Indeed, both considered leaving their programs \textit{before} finding a research group.  Meanwhile, Figure \ref{fig:paths}b) represents the path exemplified by Tabitha and Kwan, beginning with their strong research backgrounds that made them confident in the research they wanted to do in graduate school.  Both students found a research advisor before they arrived in graduate school, primarily based on their research topic.  However, upon working in their research groups during their first year, they recognized that their labs did not provide a working environment conducive to their success and considered leaving.  

Many other students' experiences are captured in the model as well, as summarized in Figure \ref{fig:paths}. For example, Elena, Nina, and Alex are illustrative of the path shown in Figure \ref{fig:paths}c).  Their undergraduate research played a defining role in determining the research they wanted to pursue in graduate school.  And although this analysis primarily focused on the role of research interests in students' search for a research group, many students in our sample were clearly cognizant of cultural factors in their choice of research group; for Elena, Nina, and Alex, their undergraduate institutions and research experiences gave them strong connections to their prospective groups.  Elena and Nina had already worked on the same collaborations as their eventual graduate advisors, and Alex's undergraduate advisor had previously worked with his prospective graduate advisor.  Thus, this subset of students had insight into both the research and cultural aspects of their research groups early on in their search process, and they joined their labs with high confidence in having a good fit.

Another notable pathway through the model is illustrated by Eric, Cole, and Kali in Figure \ref{fig:paths}d).  We referenced Eric and Cole in Section \ref{subsec:interests} as having a broad set of interests coming into graduate school, but who joined a research group early anyway.  Kali was also discussed in this section, indicating that she wanted to ``explore'' in her first year of graduate school.  Thus, none of these students applied to their schools with the intention of working with a specific professor.  Eric and Cole joined groups near the beginning of their graduate programs anyway, figuring that if it did not work out they would have time to switch. In effect, they were creating their own ``rotation systems.'' Meanwhile, Kali's program already had an institutionalized rotation program, which allowed her to try a lab and switch.  She credited the rotation system with helping her narrow her interests, saying ``I don't regret any of the rotations that I had done, because I know now that I don't want to do that [type of research]... I wouldn't have changed it.''  We did not include Kali in our count of students who switched groups since it was a mandatory aspect of her graduate program, but we include her experience here since it most closely resembles the path shown in Figure \ref{fig:paths}d).  Indeed, all three students found their first labs to be poor fits, predominantly due to the research.  Therefore, they left and found new groups by the end of their first year summer.

Lastly, in Figure \ref{fig:paths}e) we highlight several students who took a similar path as Brianna and Carmen but who never reported feeling isolation in their programs.  Rather, they were able to navigate the search process in spite of the difficulties they experienced.  Isaac and Ash had both some done undergraduate research but remained unsure of what exactly they wanted to do in graduate school.  Both reported that having to balance coursework and research was a major issue during their first year and hindered their search, and both indicated that their departments did not provide much support in helping facilitate their group search.  However, they never indicated that they were considering leaving their programs, and navigated their difficulties to find research groups by the end of the first year.  One important factor that undoubtedly contributed to this difference in outcomes was Brianna and Carmen's status as first-generation students.  They cited this several times as a major reason for their struggle adjusting to graduate school overall.  Difficulties navigating other aspects of graduate school while also worrying about finding a group undoubtedly exacerbated their negative feelings.  We suspect that other factors contributed to the different outcomes as well, including departmental culture and the presence or absence of a strong graduate student cohort, but we did not explicitly investigate these elements in this work.

\textbf{Implications for departments}  If departments are able to better identify students that are most likely to leave, they may be better equipped to provide targeted support.  Thus, one significant contribution of this work is its potential to help physics departments reduce attrition and increase the number of students who finish a PhD.  

On one hand, not all students enter graduate school knowing what topic they want to research, and may not have had as many opportunities to develop their interests as their peers.  Having broad interests in physics is not a bad thing, and is actually a characteristic that many professional physicists claim to possess \cite{verostek2023beyond}.  Plenty of students likely became attracted to physics in the first place because it allowed them to study all different aspects of the universe. However, physics graduate programs are generally not set up to support students looking to explore their research interests.  Some addressed this by joining a group quickly, creating their own ``pseudo-rotation.''  But other students might not know this is an option, or that being part of a lab community during their first semester might be beneficial.  From a funding perspective, students may be most free to explore during their first year when they are paid by the department; yet this is also the period when coursework makes exploration most difficult.  

Departments should provide clear expectations regarding when and how students should engage in the group search process.  Based on this work and our prior research \cite{verostek2023physics}, there is a compelling case that taking part in research activities in a low-commitment way early in the graduate experience is predominantly beneficial for students.  Yet official program requirements for the departments in this study generally did not indicate whether trying to engage in research during the first year was preferable, or even allowed.  

Meanwhile, handbooks often have clear expectations for passing classes and qualifying exams during the first year.  Among the schools in our sample, course requirements were the most clearly defined in all of the handbooks.  The first-year graduate curriculum in physics is highly standardized, and the presence of classes in the first year is constant across physics graduate programs.  Indeed, passing classes is often a core requirement for students to maintain good standing in their graduate programs.  It is therefore unsurprising that students who come in relying on formal departmental guidance to navigate the first year may deliberately opt to delay looking for a group in favor of passing classes.  However, this systematically disadvantages the 25\% of physics graduate students who identify as first-generation college students, as this group is less likely to reach out for help on their own  \cite{ekmekcioglu2023navigating, nsf2022earned}.  Lacking guidance to the contrary, students may perceive experiences such as being denied access to a group due to funding seem like personal failings, contributing to feelings of isolation.    

In stark contrast, we also observed individuals in our sample who had strong preferences for research before arriving to graduate school.  Several based their decisions to join a group almost entirely on research interest, which increased their risk for joining a lab with a poor working environment.  Making evaluation of the group's culture an explicit expectation during the first year might encourage such students to think about this aspect of the group search more carefully.  Moreover, in accordance with APS Bridge recommendations \cite{aps_bridge}, departments should engage in progress monitoring to give students feedback on how they are navigating the group search.  As exemplified by Tabitha and Kwan, students who appear to be navigating graduate school well on the surface might inadvertently be placing themselves in a lab that does not provide them a high chance of succeeding.  Indeed, Tabitha believed her situation was avoidable, and indicated that she would have done things differently if someone had encouraged her to talk to her advisor's former students.  

Meanwhile, it may also be  unclear as to which students are struggling with issues such as reaching out to advisors, getting funding, and beginning to feel isolated in their programs.  Providing students with a space to discuss their progress with a member of the department could help students avoid the pathways that lead to a higher risk of leaving.  One way to facilitate these meetings could be through development of an individual development plan (IDP) tool for physics graduate students. IDPs are tools designed to help students explicate their career goals and describe in detail how they plan to meet them \cite{bosch2013building, vanderford2018use, tsai2018optimizing}. Hence, IDPs serve as a way to help students know what they should be doing at each step of their doctoral process in order to achieve their goals, and may help facilitate conversation regarding progress through the program. 

Our results highlight the diversity of student experiences leading up to their enrollment in graduate school, and indicate the need for individualized advising and support to reflect this diversity.  Graduate programs cannot treat support for students monolithically, as existing structures often do. For instance, one of the primary ways that physics departments help students find a research group is through mandatory research seminars \cite{artiles2023doctoral}.  Yet research seminars only facilitate interaction in a classroom space rather than a lab space, which limits students' ability to see what the day-to-day work is like, as well as their chances to meet graduate students and postdocs.  These connections are vital for students who are in the beginning stages of looking for a group.  Meanwhile, seminars are less likely to benefit students with narrowly defined interests, or who have already committed to a group.  Their time might be better spent working with their new group trying to gauge if they fit well in the lab.

The significant time commitment to coursework and teaching responsibilities during the first year and lack of structure to meet and begin working in prospective groups place the onus on students to overcome these obstacles.  In our previous work \cite{verostek2023physics}, we suggested several structures to help students navigate these concerns.  These included formal programs to give students a trial period to work in a group during their first year, and explicitly integrating the search for a group into the graduate curriculum.

In addition to helping students figure out what research they want to pursue, explicitly creating a trial period for participating in research groups during the first year may benefit students with narrower interests as well. Switching research groups was a pivotal event for several students in our sample.  For Tabitha and Kwan, being able to switch groups kept them in their programs.  For students like Eric and Cole, knowing that they could switch groups later gave them confidence to join a group early just to try it out.  Yet many students find leaving a group to be difficult \cite{lovitts2002leaving}.  For example, they may feel like leaving a group will set them behind on their PhD timeline, or may not realize that changing advisors is an acceptable thing to do.  A structure like a rotation program would build in the practice of switching groups; however, in the absence of such a program departments should work to normalize the practice of switching research groups.  This could include guidance on etiquette for approaching their advisor with thoughts of switching groups, and how to foster communication between the student, advisor, and the director of graduate studies.

\textbf{Limitations and future research} Interviewing students during their first and second years of physics graduate school allowed us to obtain deeper insights into their thoughts ``in the moment'' while looking for a group.  However, in doing so we sacrificed some insight into these students' outcomes later on in their graduate careers.  The grey text in Figure \ref{fig:discussion} represents an area of the model that we were unable to investigate in this paper, but that we suspect to exist.  We did not have enough data to investigate the outcomes for students who join a research group later on but experience a poor fit.  Realizing that a lab is a poor fit can take time, as exemplified by Tabitha and Kwan who worked in their labs for over a year before switching.  Students who join a group later in their first year might not realize that the fit is not great until late into their second year.  By then, the cost of switching might feel too great, and could leave them feeling as though they have no choice but to stay in the poor research environment or leave the program.  More research into these students' experiences would be needed to understand how they fit into our model. 

The goal of this paper from the outset was to emphasize the role of students' research interests in how they find a group.  Thus, a clear limitation of the results presented here is that they do not offer as much insight into how students evaluate group cultures and working environments.  A positive interpersonal dynamic in a research group is clearly essential for students to thrive, but how students weigh these considerations in comparison to their research interests remains unexplored within our data. Previous work in the context of biology graduate education indicates that students weighed both factors heavily when choosing a group \cite{maher2020finding}, and we see indications of that in our data as well.  However, a fuller description is needed to understand how students approach finding a group with a positive culture.

Lastly, our data does not include students who left their graduate programs.  Their perspectives on the physics graduate experience are critically important, but are often difficult to collect due to poor tracking of students once they leave their programs \cite{lovitts2002leaving}.  Although our data included students who reported a strong consideration of leaving, they ultimately stayed.  Gaining a more accurate characterization of how finding a research group impacts student retention, satisfaction, and leaving will require more work to gather data from those who did not complete the PhD.  It is therefore important to recognize that the trends observed in this study may not adequately represent these students \cite{hammer2014confusing}.  Future research should also attend to how it is experienced across demographic groups and within different institutional contexts, as our study was limited in these regards. Our sample was diverse in many ways, but our protocol did not explicitly probe how students felt their identities impacted their group search.  Moreover, this analysis did not systematically examine differences across race, gender, or international status regarding time to join groups and development of research interests.  And although first-generation status came up naturally during our study played a significant role in several students' narratives, this was not our central focus.  Future research must attend to how physics students' identities shape the first year of the doctoral experience in order to provide more holistic and equitable support for all students.

\begin{acknowledgements}
We thank the graduate students who participated in this study. We hope their stories contribute to ongoing improvements to doctoral education in physics and beyond. We also wish to acknowledge Christian Cammarota for help with inter-rater reliability, Diana Sachmpazidi for providing valuable feedback, and Tom and Keena Wolff for graphic design support.  This work is supported by NSF Award HRD-1834516.
\end{acknowledgements}

\clearpage

\bibliography{main.bib}% Produces the bibliography via BibTeX.

\end{document}